\documentclass[%
 reprint,
 amsmath,amssymb,
 aps,
]{revtex4-2}

\usepackage{color}
\usepackage{graphicx}
\usepackage{float}
\usepackage{dcolumn}
\usepackage{bm}
\usepackage{hyperref}
\hypersetup{colorlinks = true,
            linkcolor = blue,
            anchorcolor = blue,
            citecolor = blue,
            filecolor = blue,
            urlcolor = blue
            }
\usepackage{tikz,xcolor,hyperref}
\definecolor{lime}{HTML}{A6CE39}
\DeclareRobustCommand{\orcidicon}{%
    \begin{tikzpicture}
    \draw[lime, fill=lime] (0,0) 
    circle [radius=0.16] 
    node[white] {{\fontfamily{qag}\selectfont \tiny ID}};    \draw[white, fill=white] (-0.0625,0.095) 
    circle [radius=0.007];    \end{tikzpicture}
    \hspace{-2mm}}
    \foreach \x in {A, ..., Z}{%
    \expandafter\xdef\csname orcid\x\endcsname{\noexpand\href{https://orcid.org/\csname orcidauthor\x\endcsname}{\noexpand\orcidicon}}
}

\begin{document}

\preprint{APS/123-QED}

\title{Mock data study for next-generation ground-based detectors: \\The performance loss of matched filtering due to correlated confusion noise}

\author{Shichao Wu\orcidA{}}
\author{Alexander H. Nitz\orcidB{}}
\email{alex.nitz@aei.mpg.de} 

\affiliation{%
 Max-Planck-Institut f{\"u}r Gravitationsphysik (Albert-Einstein-Institut), D-30167 Hannover, Germany\\
 Leibniz Universit{\"a}t Hannover, D-30167 Hannover, Germany
}%

\date{\today}

\begin{abstract}

The next-generation (3G/XG) ground-based gravitational-wave (GW) detectors such as Einstein Telescope (ET) and Cosmic Explorer (CE) will begin observing in the next decade. Due to the extremely high sensitivity of these detectors, the majority of stellar-mass compact-binary mergers in the entire Universe will be observed. It is also expected that 3G detectors will have significant sensitivity down to 2-7 Hz; the observed duration of binary neutron star signals could increase to several hours or days. The abundance and duration of signals will cause them to overlap in time, which may form a confusion noise that could affect the detection of individual GW sources when using naive matched filtering; matched filtering is only optimal for stationary Gaussian noise. We create mock data for CE and ET using the latest population models informed by the \texttt{GWTC-3} catalog and investigate the performance loss of matched filtering due to overlapping signals. We find the performance loss mainly comes from a deviation in the noise's measured amplitude spectral density. The redshift reach of CE (ET) can be reduced by 15\%-38\% (8\%-21\%) depending on the merger rate estimate. The direct contribution of confusion noise to the total signal-to-noise ratio is generally negligible compared to the contribution from instrumental noise. We also find that correlated confusion noise has a negligible effect on the quadrature summation rule of network SNR for ET, but might reduce the network SNR of high detector-frame mass signals for detector networks including CE if no mitigation is applied. For ET, the null stream can mitigate the astrophysical foreground. For CE, we demonstrate that a computationally efficient, straightforward single-detector signal subtraction method suppresses the total noise to almost the instrument noise level; this will allow for near-optimal searches. 
\end{abstract}

\maketitle

\section{\label{sec:intro}Introduction\protect\\}

It has been seven years since Advanced LIGO detected the first gravitational-wave (GW) event GW150914 \citep{LIGOScientific:2016aoc} on September 14th, 2015. During these seven years, Advanced LIGO and Advanced Virgo have continuously moved toward their design sensitivities \citep{Smith:2009bx,Accadia:2011zzc,Virgo:2019juy,McCuller:2020yhw}, and have performed three observation runs (O1, O2, and O3). Nearly 100 GW signals from compact binary coalescence (CBC) have been observed, of which more than 90 are from binary black hole mergers (BBH) \citep{LIGOScientific:2021djp,Nitz:2021zwj}, two come from binary neutron star inspiral (BNS) \citep{LIGOScientific:2017vwq,LIGOScientific:2020aai}, and two black hole neutron star mergers (NSBH) \citep{LIGOScientific:2021qlt} with relatively high significance. With the increasing number of observations, we know more about the population properties of these CBC sources \citep{LIGOScientific:2018jsj,LIGOScientific:2020kqk,LIGOScientific:2021psn}. KAGRA, a GW observatory in Japan that is under active development \citep{KAGRA:2022fgc}, conducted a joint observation with GEO600 \citep{Willke:2002bs} in Germany at the end of O3 \citep{LIGOScientific:2022myk}. Before 2030, a third LIGO detector, LIGO-India, is expected to be operating \citep{Saleem:2021iwi}, and all these detectors will be further upgraded toward the A+ \citep{Miller:2014kma,McCuller:2020yhw,Cahillane:2022pqm}, AdV+ \citep{Bersanetti:2021axy} and KAGRA+ \citep{Michimura:2020xnj} configurations. At that time, there will be five kilometer-scale observatories \citep{KAGRA:2013rdx}.

After 2030, these detectors will be joined by more advanced next-generation (3G/XG) GW detectors, such as Einstein Telescope (ET) \citep{Hild:2009ns,Punturo:2010zz,Punturo:2010zza,Hild:2010id,Maggiore:2019uih,DiPace:2022uzn} in Europe and Cosmic Explorer (CE) \citep{LIGOScientific:2016wof,Reitze:2019iox,Evans:2021gyd} in the United States. There will also be space-borne GW detectors, such as LISA \citep{LISA:2017pwj,Baker:2019nia} and either Taiji \citep{Gong:2011zzd,Hu:2017mde,Wu:2021kgt,yue2021china,luo2021taiji} or TianQin \citep{TianQin:2015yph,Hu:2017yoc,TianQin:2020hid,Luo:2020bls}. For the 3G detectors on the ground, the low-frequency sensitivity will be improved to enable observation from 2-7 Hz, down from the 10-20 Hz of second-generation (2G) detectors. GW signals will stay in the detectors' sensitive frequency band for hours or even days \citep{Zhu:2020ffa,Zhu:2021ram}. The higher sensitivity of 3G observatories means a significantly higher detection rate ($\mathcal{O}\left(10^{5}\right)$ BNS mergers per year); the increased detection results in numerous signals overlapping in time. This problem is in fact more serious in space-borne GW detectors, where $\mathcal{O}\left(10^{8}\right)$ white dwarf binaries from the Milky Way and nearby galaxies produce a GW foreground noise \citep{Nelemans:2001hp,Timpano:2005gm,Crowder:2006eu,Digman:2022jmp}. For stationary and Gaussian noise, the matched filter is the optimal linear filter that maximizes the signal-to-noise (SNR) \citep{turin1960introduction}. For the 2G GW detectors, the instrumental noise can at most times be assumed to be stationary and Gaussian, with notable additive nonstationary ``glitches'' \citep{Zevin:2016qwy,Cabero:2019orq,LIGO:2021ppb}; methods based on matched filtering are widely used in the search of CBC signals \citep{Owen:1998dk,Allen:2005fk,Usman:2015kfa}.

If in sufficient abundance, numerous overlapping signals can form another kind of noise in addition to the detector noise, that is, confusion noise. The authors of \citep{Regimbau:2009rk} first proposed that Einstein Telescope might be affected by confusion noise.
We note that ``confusion noise'' is used to refer to a large population of unresolvable GWs; here we collectively refer to the foreground population of overlapping signals, many of which are individually resolvable, but retain sufficient numbers to behave as an additive noise source. A few years later \citep{Regimbau:2012ir} launched the first mock data challenge for Einstein Telescope; they simulated binary neutron star (BNS) sources according to a state-of-art population model at that time, and used the \texttt{ihope} pipeline \citep{Brown:2005zs,Allen:2005fk,Babak:2012zx} based on matched filtering together with a null stream method (a technique to cancel the GW strain in the data for better detector noise estimation) \citep{Guersel:1989th,Wen:2005ui,Ajith:2006qk,Chatterji:2006nh,Wen:2007pj,Freise:2008dk,Goldstein:2017qub,Dupree:2019jqn,Wong:2021eun,Goncharov:2022dgl,Janssens:2022cty} to analyze the simulated data. They found no significant impact of confusion noise in searching with data from Einstein Telescope. A later analysis produced consistent results \citep{Meacher:2015rex}. These early mock data challenges were conducted before significant constraints on the CBC population were available. Recently, several groups have begun to study the influence of overlapping signals on the parameter estimation for CBC sources \citep{Samajdar:2021egv,Pizzati:2021apa,Smith:2021bqc,Relton:2021cax,Himemoto:2021ukb,Antonelli:2021vwg}. There is also a search study for overlapping signals in 2G cases \citep{Relton:2022whr}.

Currently, there are a few methods to improve signal detection in the signal overlapping case. The ``null stream'' is a method of cancelling astrophysical sources in detector data to achieve better power spectral density (PSD) estimation \citep{Guersel:1989th,Wen:2005ui,Ajith:2006qk,Chatterji:2006nh,Wen:2007pj,Freise:2008dk,Goldstein:2017qub,Dupree:2019jqn,Wong:2021eun,Goncharov:2022dgl,Janssens:2022cty}. It requires (1) at least 2 coaligned (colocated) detectors \citep{Ajith:2006qk,Freise:2008dk} or 3 detectors if not colocated, (2) the PSD's shape of each detector should be the same \citep{Regimbau:2012ir}, (3) it is assumed that there are only two polarizations of GWs, which is consistent with the general theory of relativity \citep{Pang:2020pfz,Wong:2021cmp}. For distant detectors, the null stream becomes sky-dependent \citep{Wen:2007pj}, it can only cancel the GW signal incident from a specific direction at a time (because the null stream combination is dependent on the time delays). Thanks to the structure of ET (closed-loop composed of three V-shaped detectors \citep{Hild:2009ns}), the distances of the three subdetectors (E1, E2, E3) are very close, so the time delay between them can be ignored; all GW signals can be canceled by appropriately summing the data from these three subdetectors. However, these assumptions (especially the first one) might be unrealistic for CE, so we need to develop other complementary methods.
For LISA, a ``global fit'' \citep{Littenberg:2020bxy} is usually used to avoid the bias caused by signal overlapping. In this method, the number of signals in the data is also used as an unknown variable and the parameter estimation of all signals is done simultaneously. The degeneracy of parameters and convergence of the sampling in such high-dimensional parameter space are potential problems.

In our study, we create mock data for CE and ET. Unlike the previous two ET mock data challenges, we use CBC population models based on the latest \texttt{GWTC-3} observational constraints. We use these simulated data to investigate several key factors that might negatively affect the performance of matched filtering, such as the bias in PSD estimation, the bias from the SNR contribution of overlapping GW signals, and the bias caused by the confusion noise's correlation between different detectors. We perform a computationally optimized matched filtering search to demonstrate identifying and subtracting the majority of high SNR sources. We find that postsubtraction, we are able to obtain a PSD close to the design sensitivity for CE. This can be used as the first-stage foreground cleaning before more sophisticated searches.

The structure of this paper is as follows: In Sec.~\ref{sec:confusion_gen}, we introduce the CBC population models used in this paper and how to simulate the time-domain mock data of CE and ET based on these models. In Sec.~\ref{sec:biases_from_confusion}, we derive the matched filtering equations in the presence of overlapping signals. In Sec.~\ref{sec:bias_from_psd},~\ref{sec:bias_from_cross_term}, and \ref{sec:bias_from_correlated_noises}, we investigate the effects of overlapping signals on PSD estimation, cross term calculation, and network SNR calculation under correlated noise. After that, we present our signal subtraction method and results in Sec.~\ref{sec:method}. Finally, Sec.~\ref{sec:conclusions} is the summary and discussion.

\section{\label{sec:confusion_gen}Population Models and Mock Data Generation}

In this paper, we need to simulate a realistic population of the known types of CBC GW sources, such as BBH, BNS, and NSBH, and then inject their GW signals into simulated instrumental noise. We create a simulated set of sources that follows the population estimates of the \texttt{GWTC-3} catalog \citep{LIGOScientific:2021psn}. Researchers have started to use the observed mergers to constrain parametric models of the population \citep{LIGOScientific:2018jsj,LIGOScientific:2020kqk,Roulet:2020wyq,Roulet:2021hcu,Zhu:2021jbw}; the nearly one hundred current observations place some constraints on the BBH population, however, the BNS and NSBH populations remain highly uncertain due to the small number of observed sources.

\subsection{\label{sec:population_model}The population models}

Quasicircular CBC systems can be described by 15 parameters; these include intrinsic parameters such as mass and spin of the components and extrinsic parameters such as sky localization, luminosity distance, orientation angle, and polarization angle. In systems containing neutron stars (such as BNS and NSBH), there are also the tidal deformation parameters of the neutron star \citep{Flanagan:2007ix,Favata:2013rwa}. We choose the prior that sources have an isotropic distribution of viewing angle, polarization angle, and sky localization. For the remaining intrinsic parameters and the distance distribution, we base our choice for the population of BNS, NSBH, and BBH parameters on the most constrained models from the latest GW catalogs.

To specify the distribution of source-frame BBH primary mass and mass ratio, we use the results of the \textit{power-law+peak} mass distribution model (shown in Fig.~10 of \citep{LIGOScientific:2021psn}). A recent study shows that the mass distribution of black holes in NSBH systems also agrees with this distribution \citep{Zhu:2021jbw}, so we adopt the same mass distribution model in the BBH and NSBH systems. For the mass distribution of neutron stars in BNS and NSBH systems, we use the distribution of the \textit{power-law} model, shown in Fig.~7 of \citep{LIGOScientific:2021psn}. There are only four confident observations containing at least one neutron star (two BNS events, GW170817 \citep{LIGOScientific:2017vwq} and GW190425 \citep{LIGOScientific:2020aai}, and two NSBH events, GW200105 and GW200115 \citep{LIGOScientific:2021qlt}) and the LVK uses all of these events to constrain the mass distribution of neutron stars. Due to the limited number of observations, the NS's population properties are poorly constrained.

We choose the spin amplitude distribution of black holes in both BBH and NSBH systems following the distribution shown as the solid black curve in Fig.~15 of \citep{LIGOScientific:2021psn}. In this work, we use an isotropic spin orientation distribution for BBH sources, as related parameters are not highly constrained. For the NSBH and BNS systems, consistent with the current observations of low spinning neutron stars \citep{LIGOScientific:2017vwq,LIGOScientific:2020aai,LIGOScientific:2021qlt}, we assume the neutron star to be orbit-aligned and slowly spinning in NSBH systems and nonspinning in BNS systems. We ignore the tidal deformation of the neutron star due to its relatively small effect on the signal's SNR \citep{Zhu:2020ffa,Cullen:2017oaz}.

 The distribution of sources in luminosity distance or redshift and the total merger rate determines the number of signals present in the simulated data. We adopt the simulation method of \citep{Zhu:2020ffa,Regimbau:2012ir,Meacher:2015rex}; we assume that all CBC systems are generated by stellar evolution, more precisely the evolution of field binaries \citep{Bavera:2020uch}, and we ignore dynamical encounters in the dense environment \citep{Fragione:2019vgr} and primordial black holes formed in the early Universe \citep{Carr:2016drx}. Since all the CBC systems in the simulation come from stellar evolution, the redshift distribution of these CBC sources must be directly related to the star formation rate (SFR). We use the SFR in \citep{Madau:2014bja}. The coalescence rate density in the source-frame is the convolution of the star formation rate and the time delay probability distribution \citep{Zhu:2020ffa}, 
\begin{equation}
\begin{aligned}
\dot{\rho}(z)=\dot{\rho}_{0} f(z) & \propto \int_{\tau_{\min }}^{\infty} \dot{\rho}_{*}\left[z_{f}(z,\tau)\right] P(\tau) d \tau \\
& \propto \int_z^{\infty} \dot{\rho}_{*}\left(z_{f}\right) P\left[\tau\left(z,z_{f}\right)\right] \frac{d t\left(z_{f}\right)}{d z_{f}} d z_{f}, \label{eq:merger_rate_den}
\end{aligned}
\end{equation}

where $\dot{\rho}_{0}$ is the local coalescence rate density (in the unit of $\text {Mpc}^{-3} \mathrm{yr}^{-1}$), which is equivalent to a rescaling factor, and $f(z)$ is the normalized coalescence rate density, such that $f(0)=1$ and $\dot{\rho}(0)=\dot{\rho}_{0}$. The SFR $\dot{\rho}_{*}$ is in the unit of $M_{\odot} \mathrm{Mpc}^{-3} \mathrm{yr}^{-1}$. $P(\tau)$ is the probability distribution of the time delay $\tau$. The time delay $\tau=t(z)-t\left(z_{f}\right)$ refers to the total time from the formation of the binary progenitor system (when redshift is $z_{f}$) to the merger of the compact binary system due to GW emission (when redshift is $z$). This delay is determined by the difference between the lookback time of $z$ and $z_{f}$, and the lookback time at redshift $z$ is defined as
\begin{equation}
t(z)=\frac{1}{H_{0}} \int_{z}^{\infty} \frac{d z}{(1+z) \sqrt{\Omega_{\Lambda}+\Omega_{\mathrm{m}}(1+z)^{3}}}, \label{eq:lookback_time}
\end{equation}

where ${H_{0}}$ is the Hubble constant, and $\Omega_{\Lambda}$ and $\Omega_{\mathrm{m}}$ are the densities of dark energy and nonrelativistic matter respectively. In this paper, we assume the standard $\Lambda \mathrm{CDM}$ cosmology \citep{Planck:2015fie}, so ${H_{0}}=67.74 \mathrm{~km} \mathrm{~s}^{-1} \mathrm{Mpc}^{-1}$, $\Omega_{\Lambda}=0.6910$, and $\Omega_{\mathrm{m}}=0.3075$. In Eq.~(\ref{eq:merger_rate_den}), we convert the integral variable from the time delay $\tau$ to the redshift $z$ for convenience. At present, there are several models of time delay distribution $P(\tau)$, such as Gaussian delay model \citep{Virgili:2009ca}, log-normal delay model \citep{Wanderman:2014eza}, power-law delay model \citep{Wanderman:2014eza} and inverse delay model \citep{Belczynski:2017gds}, the first three are derived from actual observations, and the last one is suggested by the population synthesis \citep{Lipunov:1995ct,Ando:2004pc,Belczynski:2006br,OShaughnessy:2007brt,Dominik:2012kk} and used by the previous ET mock data challenges \citep{Regimbau:2012ir,Meacher:2015rex}. We also use the inverse delay model in this paper, which means $P\left(\tau \right) \propto 1 / \tau$. In order to obtain the distribution of the event rate of CBC as a function of redshift in the detector frame, we need to multiply the coalescence rate density expressed by Eq.~(\ref{eq:merger_rate_den}) by the comoving volume element $d V(z) / d z$, and then divide it by $1+z$ caused by the time dilation, so we get the following equation
\begin{equation}
\frac{d R}{d z}=\frac{\dot{\rho}_{0} f(z)}{1+z} \frac{d V(z)}{d z}, \label{eq:merger_rate}
\end{equation}

where the comoving volume element $d V(z) / d z$ is described by
\begin{equation}
\frac{d V(z)}{d z}=\frac{c}{H_{0}} \frac{4 \pi D_{\mathrm{L}}^{2}}{(1+z)^{2} \sqrt{\Omega_{\Lambda}+\Omega_{\mathrm{m}}(1+z)^{3}}}, \label{eq:comoving_volume}
\end{equation}

where $c$ is the speed of light in the vacuum, and $D_{\mathrm{L}}$ is the luminosity distance between the CBC source and the detector, which is defined as
\begin{equation}
D_{\mathrm{L}}=(1+z) \frac{c}{H_{0}} \int_{0}^{z} \frac{d z}{\sqrt{\Omega_{\Lambda}+\Omega_{\mathrm{m}}(1+z)^{3}}}. \label{eq:luminosity_distance}
\end{equation}

\begin{figure*}
    \centering
	\includegraphics[scale=1.2]{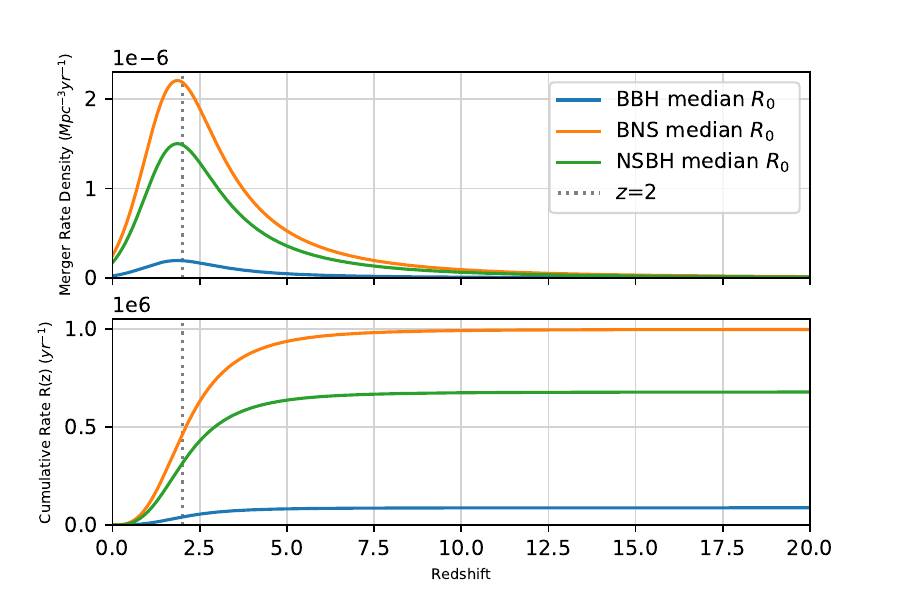}
    \caption{The merger rate density and the cumulative merger rate for the median local merger rate case are shown separately for BNS (orange), NSBH (green), and BBH (blue) sources. The results are rescaled according to the local merger rate densities (median value) constrained by \texttt{GWTC-3}. The upper panel shows the merger rate densities according to Eq.~(\ref{eq:merger_rate_den}). Note that all distributions peak around $z=2$ (the gray vertical dotted line), and then decrease monotonically. The lower panel shows the corresponding cumulative merger rates, which are calculated according to Eq.~(\ref{eq:merger_rate}). It can be seen that these curves become steepest near $z=2$, and there is almost no increase after $z=10$.}
    \label{fig:median_merger_rate}
\end{figure*}

For the local merger rate of each CBC type, we choose the rates based on LVK's population paper of \texttt{GWTC-3} \citep{LIGOScientific:2021psn} and the public presentation~\footnote{Webinar: The population of merging compact binaries inferred using GWs through \texttt{GWTC-3} \url{https://dcc.ligo.org/LIGO-G2102458/public}}. For BBH signals, we choose 22 Gpc$^{-3}$yr$^{-1}$ and 45 Gpc$^{-3}$yr$^{-1}$ as median and upper local merger rates respectively, 250 Gpc$^{-3}$yr$^{-1}$ and 1900 Gpc$^{-3}$yr$^{-1}$ for BNS signals, and 170 Gpc$^{-3}$yr$^{-1}$ and 320 Gpc$^{-3}$yr$^{-1}$ for NSBH signals. Note that in the latest version of \citep{LIGOScientific:2021psn}, they have changed the rate of NSBH to lower values (several months after our project started), but in this paper, we still use their original values. The coalescence rate of BBH, NSBH, and BNS in the detector frame as a function of redshift is shown for the median local merger rate case as the upper plot of Fig.~\ref{fig:median_merger_rate}. We draw the redshift (luminosity distance) of the GW signal from this distribution. When the redshift is higher than 20, there are few CBC sources generated by the stellar evolution, so we choose to simulate GW sources only up to redshift $z_{\max}=20$.

We can get the average time interval between two adjacent GW signals of the same type (the overline means the average) as,
\begin{equation}
\overline{\Delta t}=\left[\int_{0}^{z_{\max }} \frac{d R}{d z}(z) \mathrm{d} z\right]^{-1}. \label{eq:average_time_interval}
\end{equation}

For median local merger rate cases, we find the average time intervals for BBH, BNS, and NSBH are 359.4 s, 31.6 s, and 46.5 s respectively. For the cases of upper local merger rate, we find the average time intervals are 175.7 s, 4.2 s, and 24.7 s respectively.

\subsection{Mock data generation}

We create mock data for both Einstein Telescope and Cosmic Explorer. According to population models in the previous Sec.~\ref{sec:population_model}, we use either the median or upper local merger rate to simulate a population of sources and then project the GW signal to each detector, according to the equation
\begin{equation}
h^{\alpha}(t)=F_{+}^{\alpha}(\theta, \phi, \psi) h_{+}(t)+F_{\times}^{\alpha}(\theta, \phi, \psi) h_{\times}(t), \label{eq:det_strain}
\end{equation}

where $\alpha$ is the index of each detector, $F_{+}^{\alpha}(\theta, \phi, \psi)$ and $F_{\times}^{\alpha}(\theta, \phi, \psi)$ are antenna pattern functions for the two GW polarizations, which depend on the sky localization ($\theta, \phi$) and polarization angle $\psi$. These parameters are sampled from distributions defined in Section \ref{sec:population_model}. We generate and save each type of GW signal separately, then add them into the simulated Gaussian detector noise to create two datasets; each dataset has $\sim6$ hours of data, one for the median local merger rate case, the other one for the upper local merger rate case.

We use the latest time-domain phenomenological higher-order mode model \texttt{IMRPhenomTPHM} \citep{Estelles:2021gvs} to simulate BBH and NSBH sources. As the mass ratio of BBH and NSBH systems can be large, higher-order modes will have an impact on GW waveforms \citep{London:2017bcn}. We expect the tidal deformation of neutron stars in NSBH systems is relatively weak \citep{Zhu:2020ffa,Zhu:2021ysz}, so we do not use the two existing NSBH models which include the tidal effect \citep{Thompson:2020nei,Matas:2020wab}. In addition, these two models only consider the dominant (2, $\pm$2) mode.
For BNS sources, we use the frequency-domain phenomenological model \texttt{IMRPhenomD} \citep{Khan:2015jqa} instead of the post-Newtonian waveforms \citep{Buonanno:2009zt} or \texttt{IMRPhenomPv2\_NRTidalv2} \citep{Dietrich:2019kaq}. Next-generation GW detectors may observe high-redshift BNS signals, so the merger and postmerger part of the redshifted signal might be within the detector's sensitive frequency band \citep{Zhu:2021ram}. However, accurate models of merger and postmerger waveform of BNS are still under development \citep{Breschi:2019srl,Breschi:2022xnc, Breschi:2022ens}, we use the merger and ringdown parts of \texttt{IMRPhenomD} to mimic them. We also neglect the tidal deformation of neutron stars for BNS systems in this paper. In the source frame, the postmerger signal of a BNS is in the kHz frequency band \citep{Breschi:2022ens}, according to our population model most BNSs are at $z\sim2$. The SNR of a BNS postmerger  at 70 Mpc is $\sim8$ for ET, whereas the inspiral would have an SNR of $\sim1000$ (we have used \texttt{IMRPhenomPv2\_NRTidalv2} model to calculate SNR for its inspiral part); the relative SNR in the postmerger is negligible \citep{Breschi:2022ens}. If we rescale this example signal to $z\sim2$, we find an inspiral SNR is $\sim12$, while postmerger is $\sim0.2$. For the purposes of detection, more than 99.9\% of the SNR is recovered without the postmerger signal. We do not expect the presence of a post-merger signal to bias PSD estimation due to the negligible relative SNR which is contained within a shorter duration; most of the data will all be free of contamination by postmerger signals. We neglect tidal effects as they will be completely subdominant for the parts of the signal where there are a significant number of overlapping sources.

\subsubsection{\label{sec:ET}Einstein Telescope}

Einstein Telescope (ET) \citep{Hild:2009ns,Punturo:2010zz,Punturo:2010zza,Hild:2010id,Maggiore:2019uih,DiPace:2022uzn} is the European plan for the next-generation observatory following Advanced Virgo \citep{Accadia:2011zzc,LIGO:2018pgl,Bersanetti:2021axy}. ET consists of three V-shaped subdetectors (called E1, E2, and E3); they overlap with each other to form an equilateral triangle, with an arm length of 10 km. In order to achieve the required sensitivity at both low frequency and high frequency, each V-shaped subdetector uses a ``xylophone'' configuration, i.e. there is one interferometer for low frequencies and another for high frequencies \citep{Hild:2009ns,Hild:2010id}. This design is intended to bring the sensitive frequency band of ET down to $\sim2$ Hz. At present, the site selection of ET is still under evaluation. We use the fiducial coordinates and orientation defined in the \texttt{LALSuite} \citep{lalsuite} and assume the ET design sensitivity curve \citep{Hild:2009ns,Hild:2010id}. Each of ET's subdetectors has a different antenna pattern \citep{Regimbau:2012ir}; the V-shaped detector has a sensitivity loss of $1/\sin\frac{\pi}{3}$ compared to an equivalent length L-shaped detector (caused by the opening angle is $\frac{\pi}{3}$, not $\frac{\pi}{2}$) \citep{Hild:2010id,Regimbau:2012ir}. We neglect the effect of the Earth's rotation in this paper. In our simulations, we set the low-frequency cutoff of the ET dataset to 2 Hz.

\subsubsection{\label{sec:CE}Cosmic Explorer}

\begin{figure*}
    \centering
	\includegraphics[scale=1.2]{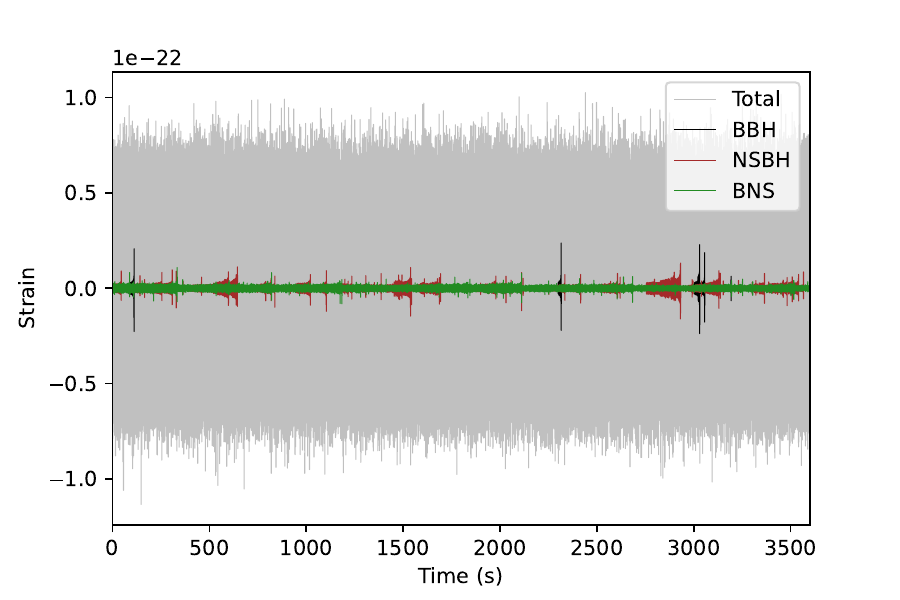}
    \caption{An hour of simulated data for CE assuming median local merger rates and corresponding population models. The gray line represents the sum of all GW signals and detector noise. The black, brown, and green lines represent the injected BBH, NSBH, and BNS signals, respectively. Because of the difference in the event rate and the signal duration, we can see that BNS signals overlap in time, with no gaps between signals, NSBH signals also have large overlaps, while many BBH signals remain isolated.}
    \label{fig:strain_median_CE}
\end{figure*}

 Cosmic Explorer (CE) is a proposed next-generation observatory in USA \citep{LIGOScientific:2016wof,Reitze:2019iox,Evans:2021gyd}. Compared with Einstein Telescope, CE adopts a more conservative design. CE uses the L-shaped configuration of the current-generation observatories, but with an arm length of 40 km. The design enables sensitivity to frequencies down to $\sim5-7$ Hz. CE and ET have their own advantages, and they can form a 3G detector network to further improve the overall signal detection capabilities \citep{Zhao:2017cbb,Zhu:2020ffa,Zhu:2021ram,Borhanian:2022czq}. Similar to ET, the location of CE has not yet been determined. In this paper, we place one CE at the position of the LIGO-Hanford observatory \citep{Smith:2009bx}. We use the latest CE design sensitivity \citep{Evans:2021gyd} to recolor the stationary, Gaussian, and whitened noise, and set the low-frequency cutoff of the CE dataset to 5 Hz. We show an hour of simulated data in Fig.~\ref{fig:strain_median_CE}. Note that while BBH signals are still largely separated in time, BNS signals overlap in time. Due to the improved low-frequency sensitivity, each BNS signal in the detector sensitivity band will last for hours or even days \citep{Regimbau:2012ir,Meacher:2015rex,Zhu:2021ram}. However, we note that the majority of signals will remain distinguishable, because their time-frequency evolution is different from each other.

\section{\label{sec:biases_from_confusion}The Biases of Matched Filtering Caused by Confusion Noise}

Assuming additive Gaussian and stationary noise, and for a known signal, the matched filter is the optimal linear filter that can maximize SNR \citep{turin1960introduction}. As a consequence, matched filtering is widely used as the basis of modeled searches for CBC sources \citep{Owen:1998dk,Allen:2005fk,Usman:2015kfa}.
In this section, first we briefly review the basic principles of the matched filtering method, and then we study whether these overlapping GW signals will have negative effects on the performance of matched filtering. 

For GW signals from the compact binary coalescence, there are state-of-art methods to numerically simulate the GW signals predicted by general relativity \citep{Baumgarte:2010ndz,Baumgarte:2021skc}. However, in order to meet the speed requirements of data analysis, there are several kinds of approximants, including post-Newtonian approximation models \citep{Buonanno:2009zt}, phenomenological models \citep{Khan:2015jqa,Hannam:2013oca,Pratten:2020ceb,Estelles:2021gvs}, effective-one-body numerical relativity models \citep{Buonanno:2000ef,Ossokine:2020kjp}, and surrogate models \citep{Field:2013cfa,Rifat:2019ltp}; these models may be compared and calibrated with the results of numerical relativity simulations. 

 In order to understand the behavior of the matched filter, we first define the scalar or inner product in the frequency domain as below
\begin{equation}
\left\langle a \mid b\right\rangle=4 \Re \int_{f_{\text {min}}}^{f_{\text {max}}} \frac{\tilde{a}(f) \tilde{b}^{*}(f)}{S_{h}^{tot}(f)} \mathrm{~d} f,
\label{eq:inner_product}
\end{equation}

$a$ and $b$ are two time series, the $\tilde{a}(f)$ is the Fourier transform of $a$, the $\tilde{b}^{*}(f)$ is the complex conjugate of the Fourier transform of $b$, the $f_{\text {min}}$ and the $f_{\text {max}}$ are the low frequency cutoff and the high frequency cutoff respectively. The $S_{h}^{tot}(f)$ is the one-sided PSD of the data $s(t)$ (here we include the contribution of all overlapping signals, not just the $S_{n}(f)$ computed from the pure detector noise), defined by
\begin{equation}
\left\langle\tilde{s}(f) \tilde{s}^{*}\left(f^{\prime}\right)\right\rangle=\frac{1}{2} S_{h}^{tot}(f) \delta\left(f-f^{\prime}\right),
\label{eq:onesided_psd}
\end{equation}

$\tilde{s}(f)$ is the Fourier-transformed detector data, defined by $\tilde{s}(f)=\int_{-\infty}^{+\infty} s(t) \mathrm{e}^{-2 \pi \mathrm{i} t f} \mathrm{~d} t$. The ``$\langle\cdot\cdot\rangle$'' here means the average of many noise realizations, not the same as the inner product ``$\left \langle \cdot | \cdot \right \rangle $''. In the general case, the time-domain observational data $s(t)$ from a GW detector is
\begin{equation}
s(t)=n(t)+\sum_{j} h_{j}(t)=n(t)+\sum_{j \neq k}h_{j}(t)+h_{k}(t),
\label{eq:strain_sum}
\end{equation}

which is the linear summation of detector noise $n(t)$ and all GW signals $\sum_{j} h_{j}(t)$ (among them $h_{k}(t)$ is the signal of interest), $j$ is the number of GW signals in this data, and it is unknown before data analysis. If there is only one GW signal $h_{k}(t)$ in the data, then $\sum_{j \neq k}h_{j}(t)$ in the Eq.~(\ref{eq:strain_sum}) disappears.

If we have a template $h(t,\Theta)$ where the merger is at $t_{c}=0$ and $\Theta$ represents all parameters of the waveform, the template with arbitrary merger time $t_{c}$ is $\tilde{h}(f,\Theta) e^{2 \pi i f t_{c}}$. By substitution into Eq.~(\ref{eq:inner_product}), we can define
\begin{equation}
\left\langle s \mid h\right\rangle(t_{c})=4 \Re \int_{f_{\text {min}}}^{f_{\text {max}}} \frac{\tilde{s}(f) \tilde{h}^{*}(f)}{S_{h}^{tot}(f)} \mathrm{e}^{2 \pi \mathrm{i} f t_{c}} \mathrm{~d} f,
\label{eq:mf_snr}
\end{equation}
which is just the inverse Fourier transform of the inner product of $s$ and $h$. An efficient search for the signal with unknown time can be efficiently done with a Fourier transform. According to Eq.~(\ref{eq:strain_sum}), we can rewrite Eq.~(\ref{eq:mf_snr}) as
\begin{equation}
\left\langle s \mid h\right\rangle=\left\langle n \mid h\right\rangle+\sum_{j \neq k}\left\langle h_{j} \mid h\right\rangle+\left\langle h_{k} \mid h\right\rangle.
\label{eq:cross_term}
\end{equation}

Note that the inner product has contributions from the detector noise $n(t)$, confusion noise $\sum_{j \neq k}h_{j}(t)$ made by overlapping signals, and the particular GW signal $h_{k}(t)$ that we are interested in.

According to Eq.~(\ref{eq:det_strain}), the detector strain caused by the GW signal is a linear combination of two GW polarizations, and the combination coefficients are determined by the antenna response function of the detector, which depends on the sky location and polarization angle of the signal. Together with the source's luminosity distance, chirp mass, and inclination angle, these factors affect the overall amplitude of the signal's strain. For typical CBC matched filtering searches \citep{Maggiore:2007ulw,Allen:2005fk,Usman:2015kfa} (quasicircular, nonprecessing, and only the dominant (2, $\pm$2) mode), it is usually assumed that the two polarizations of the signal satisfy the relationship $h_{+}=ih_{\times}$, which means that the two polarizations are related by a phase difference of $\frac{\pi}{2}$. To maximize over both an overall orbital phase and previously mentioned angles, we only need to use one of the polarizations $h_{+}$ along with the complex matched filtering SNR $\rho(t)$,
\begin{equation}
\begin{aligned}
\rho^{2}(t)&=\frac{\left\langle s \mid h_{+}\right\rangle^{2}}{\left\langle h_{+} \mid h_{+}\right\rangle}+\frac{\left\langle s \mid h_{\times}\right\rangle^{2}}{\left\langle h_{\times} \mid h_{\times}\right\rangle}=\frac{\left\langle s \mid h_{+}\right\rangle^{2}+\left\langle s \mid h_{\times}\right\rangle^{2}}{\left\langle h_{+} \mid h_{+}\right\rangle} \\
&=\frac{1}{\sigma^{2}}\left|4 \int_{f_{\text {min}}}^{f_{\text {max}}} \frac{\tilde{s}(f) \tilde{h_{+}}^{*}(f)}{S_{h}^{tot}(f)} \mathrm{e}^{2 \pi \mathrm{i} f t} \mathrm{~d} f\right|^{2},
\label{eq:complex_snr}
\end{aligned}
\end{equation}

the $\sigma^{2}$ is the variance of the noise and is also known as the optimal SNR, which is the matched filtering SNR when the template perfectly matches the signal in the absence of noise. It can be expressed as
\begin{equation}
\sigma^{2}=4 \int_{f_{\text {min}}}^{f_{\text {max}}} \frac{\left|\tilde{h_{+}}(f)\right|^{2}}{S_{h}^{tot}(f)} \mathrm{~d} f.
\label{eq:optimal_snr}
\end{equation}

In the following subsections, we examine the performance loss of matched filtering caused by the bias in the estimation of the PSD caused by numerous overlapping signals, the bias due to the cross terms between the waveform template and overlapping signals in the data, and the impact of correlated noise on a network of detectors.

\subsection{\label{sec:bias_from_psd}Power spectral density estimation}

In this subsection, we discuss the impact of overlapping signals on PSD estimation and the loss of the detector's horizon distance \citep{Allen:2005fk,Chen:2017wpg,Evans:2021gyd} due to a biased estimate of the instrumental noise. We can see in Eq.~(\ref{eq:onesided_psd}) and Eq.~(\ref{eq:strain_sum}) that if there are overlapping signals, $S_{h}^{tot}(f)$ should be higher than the instrumental-only $S_{n}(f)$. We use the median Welch estimation method \citep{welch1967use,Allen:2005fk} to calculate the PSD of our mock data. We use 512 s of data to estimate the PSD from 35 different times from each of our datasets. The reason for choosing 512 s is just following the PSD estimation in the current real searches, we also have tried other values, and their results are similar. For each 512 s segment, we use 16 s as the subsegment and 50\% overlap to calculate a PSD using the Welch averaging method. We use these 35 PSDs to calculate the mean PSD for the entire dataset as well as the 1-$\sigma$ confidence interval for each frequency.

\begin{figure*}
    \centering
	\includegraphics[scale=0.6]{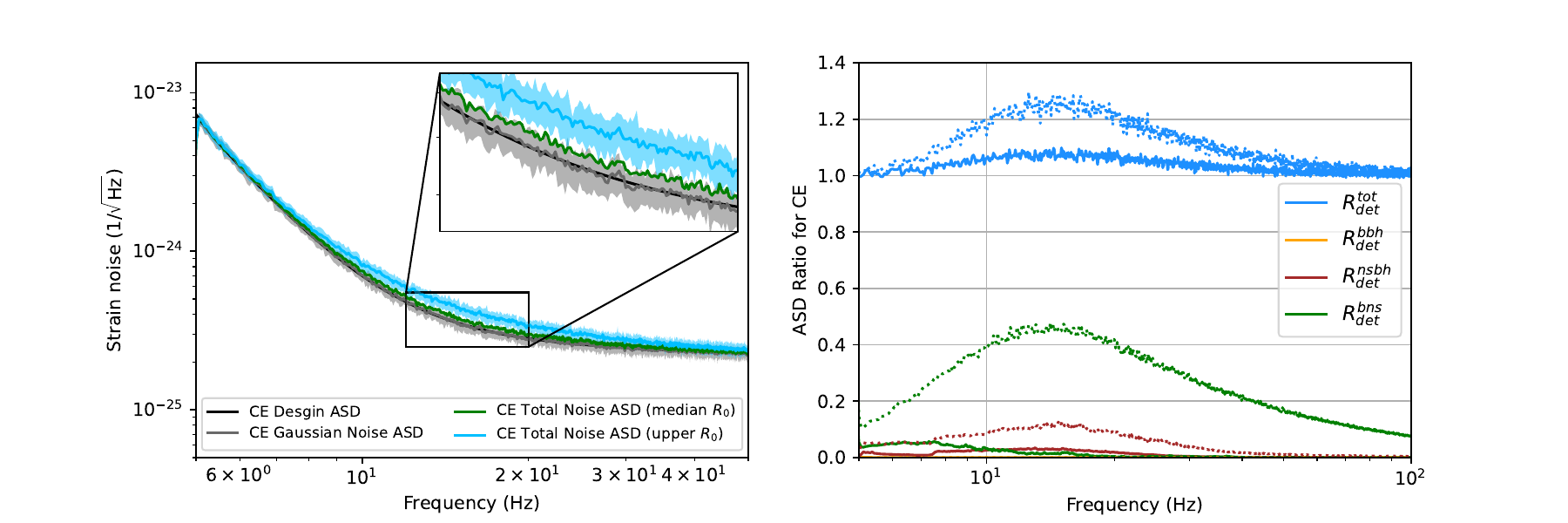}
    \caption{The different amplitude spectral densities (ASD) of CE (left) and the ratio of the estimated ASD with the signal population to data that only contains instrumental noise (right). The left panel shows the ASDs in the range of 5-50 Hz. The noise curves are shown for the ideal CE design sensitivity (black), Gaussian noise-only estimate (gray), noise including signals at the median merger rate (green), and the upper merger rate (blue). For the noise-only and upper merger rate cases, a 1-$\sigma$ confidence band is shown.
 The right panel shows the ratio of different contributions to the detector noise-only ASD. The ratio of the total ASD to the detector noise ASD (blue) is shown in addition to the contributions from BNS (green), NSBH (brown), and BBH (orange) sources for both the median (solid) and upper (dotted) merger rate cases. If there is a monochromatic GW signal, the total ratio (blue) at each frequency gives the SNR reduction factor due to the confusion noise. In the upper rate scenario, BNS sources dominate the ASD bias and form a quasistationary foreground noise. BBH sources have negligible effects on the ASD for both rate cases because their presence leaves most of the data uncontaminated. Noise estimation that uses median or median-mean Welch averaging is not significantly affected by a small number of outliers at a given frequency. The peak in the ASD bias for CE is around 15 Hz.}
    \label{fig:asd_bias_and_ratio_CE}
\end{figure*}

On the left of Fig.~\ref{fig:asd_bias_and_ratio_CE}, we show the total estimated amplitude spectral density (ASD, the square root of the PSD) of CE for the median and upper local merger rates. For comparison, we also plot the design ASD of CE and the mean ASD of simulated detector noise (without signals) for comparison. We see that the ASD of the median merger rate data is only slightly higher than the design ASD and the mean ASD of the instrumental noise, just at the boundary of the 1-$\sigma$ estimate of the instrumental noise, however, there is a significant bias for the upper merger rate dataset.

\begin{figure*}
    \centering
	\includegraphics[scale=0.6]{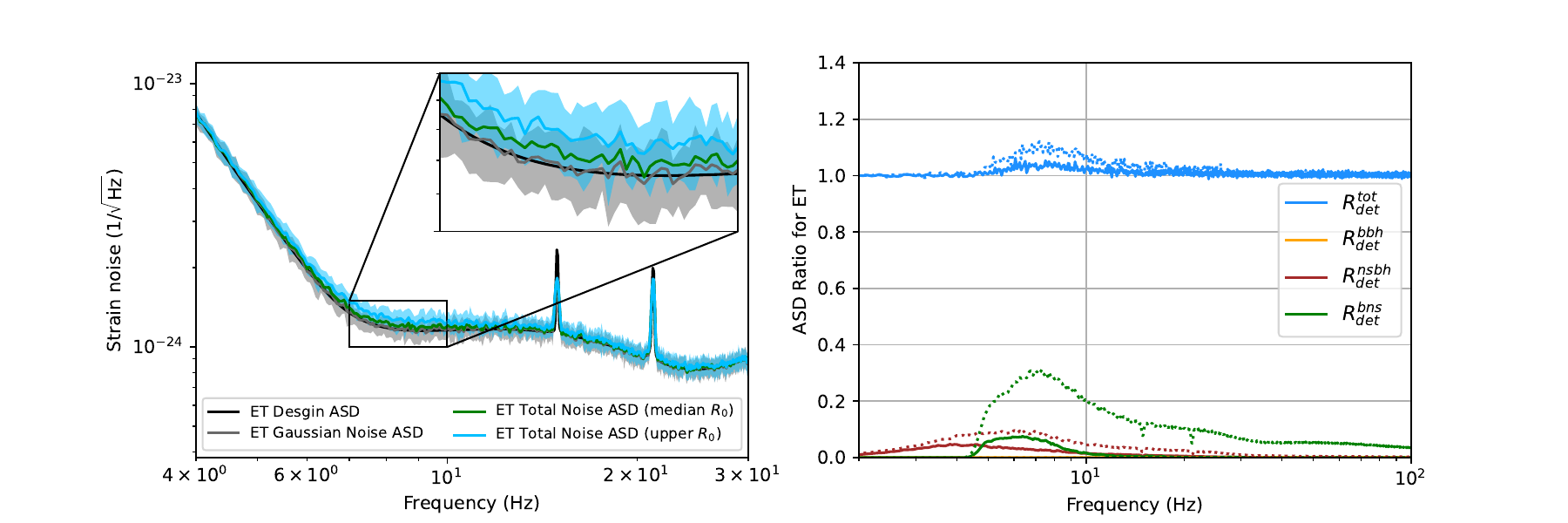}
    \caption{The different amplitude spectral densities (ASD) of ET and ASD ratios. The left panel shows the ASDs in the range of 4-30 Hz. The colors and line styles are consistent with Fig.~\ref{fig:asd_bias_and_ratio_CE}. For ET's ASD bias, the peak is around 7 Hz.}
    \label{fig:asd_bias_and_ratio_ETD}
\end{figure*}

If the value of the local merger rate is between the range used in our paper, the overall ASD will be between the green and blue lines shown in the figure. For ET, we show similar curves on the left of Fig.~\ref{fig:asd_bias_and_ratio_ETD}. It can be seen that the overall impact of overlapping signals on the ASD of ET is much smaller. Only in the case of upper local merger rate, will there be deviations noticeably above the detector noise's mean ASD between 5 Hz and 11 Hz.

To understand the contribution of each kind of source to the total ASD of CE, in the right panel of Fig.~\ref{fig:asd_bias_and_ratio_CE}, we compare the mean ASD of the simulated data only containing each type of source (BNS, NSBH, BBH) to the instrumental noise. Each ASD is plotted as deviations relative to the signal-free detector noise curve. For the median local merger rate case, we notice a deviation in the total ASD, which includes contributions from all signal classes and the instrumental noise, of up to about 10\% in the 5 Hz to 60 Hz range, and for the upper local merger rate, in this case, the deviation extends to 100 Hz, and in the range of 10 Hz to 20 Hz, the deviation can reach 20\%.

BNS mergers are the main source of bias in the measured ASD. 
For the median local merger rate case, the BNS-only deviation is mainly concentrated below 10 Hz and less than 10\%, while for the upper local merger case, the deviation is most pronounced between 10 Hz and 20 Hz, up to 40\%. We notice that the confusion noise from BNS is almost stationary, which means its PSD does not vary too much over time. The contribution of NSBH is far less than that of BNS and for the BBH datasets, regardless of the merger rate, the contribution to the total ASD is negligible.

Similarly, for ET we find the ASD deviation mainly comes from the overlapping BNS signals. The deviation is mainly concentrated between 5 Hz and 10 Hz. For the upper local merger rate case, the deviation is at most about 10\%, and for the median local merger rate, the deviation is at most about 5\%. The overall deviation is much smaller than that of CE.

\begin{figure*}
    \centering
	\includegraphics[scale=0.6]{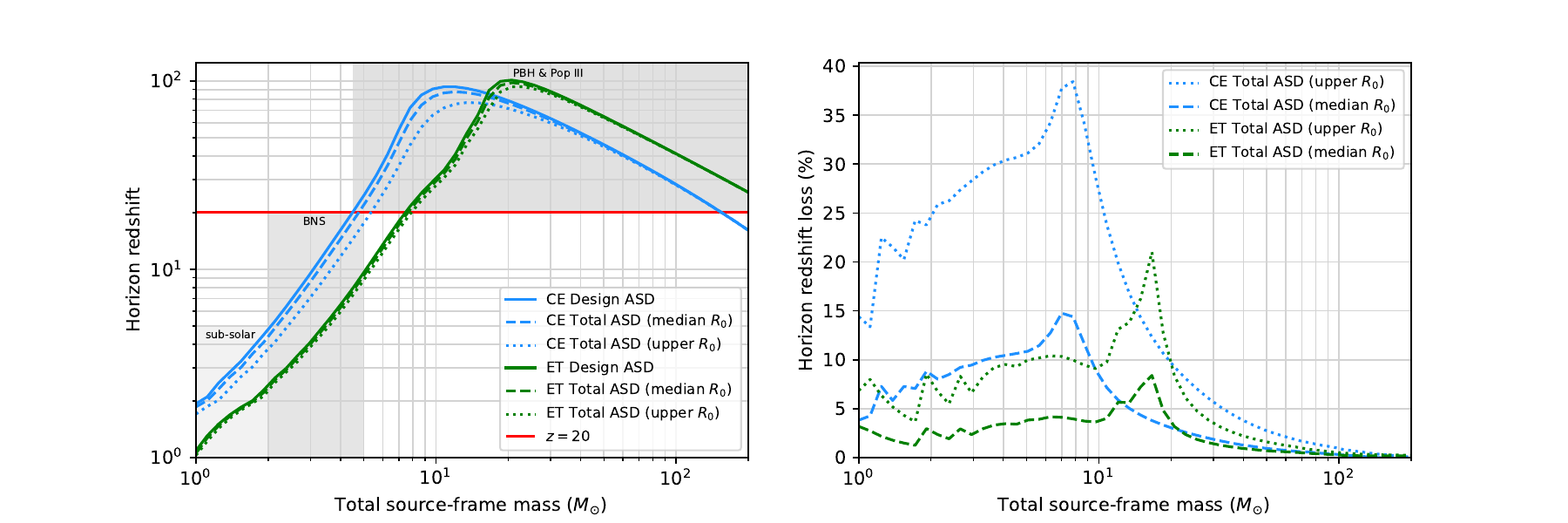}
    \caption{The horizon redshift and horizon redshift loss for different ASDs. The left plot shows the horizon redshift as a function of the source-frame total mass for the design sensitivity (solid lines) of CE (blue) and ET (green) and for the median (dashed lines) and upper (dotted lines) merger rate scenarios. Gray shaded areas roughly mark the population of subsolar compact binary, BNS, PBH, and Pop III sources. The observation of these GW sources will be affected by the confusion noise. The right plot shows the loss percentage of CE and ET's horizon redshift relative to that of the design ASD. Note we use the combined sensitivity of ET's component detectors (E1, E2, E3).}
    \label{fig:horizon_redshift_loss_ratio_CE_ET_sourceframe_totM_IMRPhenomD}
\end{figure*}

In order to study the loss of signal detection caused by biased estimation of the PSD or ASD, we calculate the horizon redshift (distance) under different conditions. The horizon redshift (distance) means the redshift (luminosity distance) of the GW source when the GW source is above the detector plane, face-on, and the optimal SNR is 8, a threshold often used to characterize the sensitivity of the detector \citep{Allen:2005fk,Chen:2017wpg,Evans:2021gyd}. The results are shown on the left side of Fig.~\ref{fig:horizon_redshift_loss_ratio_CE_ET_sourceframe_totM_IMRPhenomD}, which shows the horizon redshift for GW sources with different source-frame total masses. The confusion noise-free estimates are consistent with \citep{Evans:2021gyd}.

The existence of a large population of foreground signals, left unmitigated, would reduce the sensitivity to CBC mergers; the most extreme sensitivity loss is for sources with source-frame total mass $\sim$ 10 $M_{\odot}$. For the median local merger rate case of CE, the horizon redshift loss of CBCs with source-frame total mass less than 10 $M_{\odot}$ will be 5\% to 15\%, and for the upper merger rate case, the loss will be as high as 15\% to nearly 40\%. For ET, the loss is generally lower than CE, for the median local merger rate case, the loss is about 2\% to 7\%, and for the upper local merger case, the loss can reach 5\% to 20\%.

The loss in sensitivity may impact science at various masses. For example, sources with the total source-frame mass between 2 and 3 $M_{\odot}$ are useful to study the minimum mass of the neutron star \citep{Chatziioannou:2020msi}. Sources with a total source-frame mass between 1 and 2 $M_{\odot}$ may be primordial in origin and are the target of subsolar searches \citep{LIGOScientific:2019kan,Nitz:2021vqh}. For this kind of source, we find a 2\% to 25\% loss in the horizon redshift. For CBCs with a redshift higher than 20, it is generally considered that they might originate from Population III (Pop III) stars or primordial black hole (PBH) mergers formed in very early Universe  \citep{Koushiappas:2017kqm,Ng:2020qpk,Martinelli:2022elq,Ng:2022agi}, because these two types of sources do not follow the stellar evolution, and therefore do not follow the redshift distribution of Fig.~\ref{fig:median_merger_rate} (this distribution is only valid for field binaries). The detection of such events at high redshift can be clearly distinguished from the typical stellar-origin CBCs \citep{Ng:2020qpk,Ng:2022agi}; the presence of overlapping signals will also significantly reduce their detection efficiency.

\subsection{\label{sec:bias_from_cross_term}Matched filtering cross terms}

\begin{figure*}
    \centering
	\includegraphics[scale=0.6]{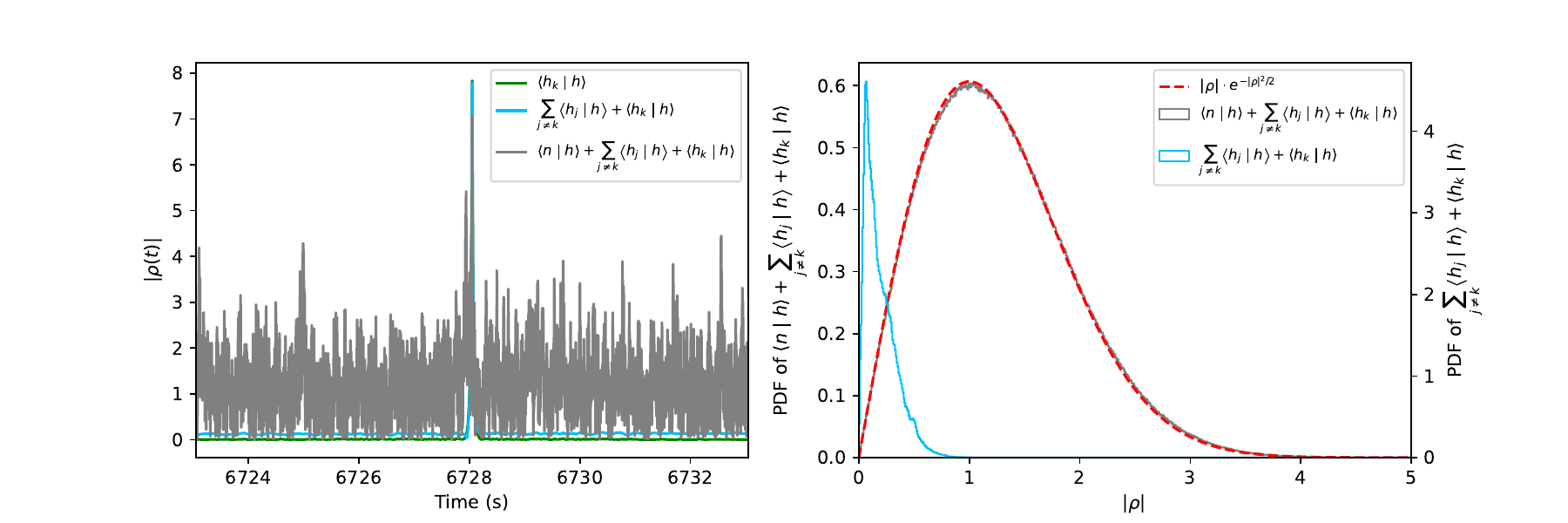}
    \caption{The absolute value of the complex SNR time series for a specific template waveform and its constituent components. The plot on the left shows the matched filtering SNR time series of an injected signal with its best matching template. Different colors represent the results of different contributions to the matched filtering SNR. The green line represents the result when there is only the injected signal (no detector noise and other signals). The blue line represents the result when other overlapping signals are also included in the data. The gray line represents the result that the data additionally contains the detector noise. Due to the fluctuation of the detector noise and confusion noise, the peak value of the gray line drops to around 7. The plot on the right shows the normalized histogram (probability density function) of the SNR time series around the time of the left panel. Even if overlapping signals are included, the total SNR time series (the gray line) is still consistent with the expected Rayleigh distribution (red dashed line), but the confusion noise itself does not follow this distribution (the blue line).}
    \label{fig:snr_timeseries_rayleigh_CE}
\end{figure*}

In this subsection, we discuss the effect that cross terms in Eq.~(\ref{eq:cross_term}) produced by overlapping signals have on the measured matched filtering SNR. In order to visually show the bias caused by this term, we show the matched filter output around
an example signal in Fig.~\ref{fig:snr_timeseries_rayleigh_CE} for the median merger rate data. For simplicity, we directly use the mass parameters of this injected signal to generate the GW template and then calculate the complex matched filtering SNR according to Eq.~(\ref{eq:complex_snr}); we use the corresponding mean PSD (the solid green one) in Sec.~\ref{sec:bias_from_psd}. To understand the impact of each component of the matched filtering SNR, we examine: (1) the data with only a specific injected signal and no detector noise, (2) the data containing all injected BNS signals but no detector noise, and (3) the data containing all injected BNS signals and the detector noise. The left side of Fig.~\ref{fig:snr_timeseries_rayleigh_CE} shows 10 s centered on the injection time. The visual inspection makes it clear that the impact of overlapping signals is negligible in this example.

We plot the probability density function (PDF) from $\sim 2800$ s of each complex SNR time series $\left| \rho(t) \right|$ in the right panel of Fig~\ref{fig:snr_timeseries_rayleigh_CE}. According to Eq.~(\ref{eq:complex_snr}), if there is only Gaussian and stationary noise of the detector in the data, then $\left\langle s \mid h_{+}\right\rangle^{2}$ and $\left\langle s \mid h_{\times}\right\rangle^{2}$ are the squares of two Gaussian variables, so $\rho^{2}(t)$ follows the chi-squared distribution with 2 degrees of freedom, and the degree of freedom is 2 because it is the sum of the squares of 2 Gaussian variables. So by definition, the modulus of $\rho(t)$ follows the Rayleigh distribution \citep{Maggiore:2007ulw}, as shown by the red dashed line on the right side of Fig.~\ref{fig:snr_timeseries_rayleigh_CE}. The total matched filtering SNR distribution only marginally deviates from the Rayleigh distribution, just slightly shifting to the direction of high SNR, this indicates the confusion noise will slightly increase the signal's SNR on average. As expected, the detector noise dominates in the 2800 s data. The SNR of the GW signal only has a peak around 8 on the right (not shown in this histogram, because of the number of bins). The PDF of the confusion noise's SNR does not follow the Rayleigh distribution expected from the Gaussian noise. The overall value is smaller than 1 and the peak value is close to 0.

\subsection{\label{sec:bias_from_correlated_noises}Correlated noise in the detector network}

In this subsection, we investigate the bias caused by the correlation of confusion noise among different detectors. For the matched filtering SNR of the detector network, we generally assume different detectors or data can be combined in quadrature as
\begin{equation}
\rho_{net} \equiv \sqrt{\sum_{i} \rho_{i}^{2}},
\label{eq:network_snr}
\end{equation}

where  $\rho_{i}$ is the matched filtering SNR of the $i$th detector in the detector network. This formula is strictly valid only when the noise of each detector is statistically independent of each other \citep{Cutler:1994ys,Finn:2000hj}. In the absence of confusion noise, if the distance between detectors is far enough, this is a reasonable assumption. Real GW searches use Eq.~(\ref{eq:network_snr}) to calculate the network SNR \citep{Usman:2015kfa}. However, for the 3G detector network, even if the instrumental noise is uncorrelated, the confusion noise composed of overlapping GW signals will still be correlated between different detectors [see Eq.~(\ref{eq:det_strain})], so there will be a correlation in the total noise, and Eq.~(\ref{eq:network_snr}) might no longer hold.

Next, we will discuss the network SNR of the detector network composed of two 3G detectors (such as two CEs) based on the method of \citep{Cutler:1994ys,Seto:2021crt}. From the Fourier transform of the time-domain strain [Eq.~(\ref{eq:strain_sum})], we can obtain the frequency-domain strain of the first detector $\mathrm{I}$
\begin{equation}
s_{\mathrm{I}}(f)=n_{\mathrm{I} D}(f)+n_{\mathrm{I} C}(f)+h_{\mathrm{I}}(f),
\label{eq:strain_sum_freq}
\end{equation}
and the detector noise $n(f)$, the confusion noise $\sum_{j \neq k}h_{j}(f)$, and the signal $h_{k}(f)$ are given as $n_{\mathrm{I} D}(f)$, $n_{\mathrm{I} C}(f)$, and $h_{\mathrm{I}}(f)$, respectively. According to Eq.~(\ref{eq:onesided_psd}), we can get the one-sided PSD for $n_{\mathrm{I} D}(f)$ and $n_{\mathrm{I} C}(f)$ as
\begin{equation}
\begin{aligned}
&\left\langle n_{\mathrm{I} D}(f) n_{\mathrm{I} D}^{*}\left(f^{\prime}\right)\right\rangle=\frac{1}{2} S_{D}(f) \delta\left(f-f^{\prime}\right), \\
&\left\langle n_{\mathrm{I} C}(f) n_{\mathrm{I} C}^{*}\left(f^{\prime}\right)\right\rangle=\frac{1}{2} S_{C}(f) \delta\left(f-f^{\prime}\right),
\end{aligned}
\label{eq:onesided_psd_det_con_I}
\end{equation}
similarly, for the second detector $\mathrm{II}$ we have the same results, just change the index from ``$\mathrm{I}$'' to ``$\mathrm{II}$.''
We assume the PSDs of these two detectors are same and the confusion noise is isotropic. According to the independence of each component, we have ($m,n=\mathrm{I},\mathrm{II}$)
\begin{equation}
\begin{aligned}
\left\langle n_{m D}(f) n_{n D}^{*}\left(f^{\prime}\right)\right\rangle=\left\langle n_{D}(f) n_{C}^{*}\left(f^{\prime}\right)\right\rangle=0,
\end{aligned}
\label{eq:onesided_psd_cross}
\end{equation}
and for the confusion noise in two detectors, we have 
\begin{equation}
\left\langle n_{m C}(f) n_{n C}^{*}\left(f^{\prime}\right)\right\rangle=\frac{1}{2} S_{C}(f) \gamma(f) \delta\left(f-f^{\prime}\right),
\label{eq:onesided_psd_det_con_I_II}
\end{equation}
where $\gamma(f)$ is the overlap reduction function (ORF) \citep{Fotopoulos:2008yq,LIGOScientific:2014sej}, which only depends on the relative position and orientation between two detectors, $-1 \leq \gamma(f) \leq 1$.
Then we can rewrite Eq.~(\ref{eq:onesided_psd_det_con_I}) and corresponding equations for the detector $\mathrm{II}$ in a matrix form, which is the noise matrix or the PSD matrix \citep{Cutler:1994ys,Seto:2021crt},
\begin{equation}
\begin{aligned}
S_{m n}(f) & =\left\langle\left[n_{m D}(f)+n_{m C}(f)\right]\left[n_{n D}(f)+n_{n C}(f)\right]^{*}\right\rangle \\
&=\left[S_{D}(f)+S_{C}(f)\right]\left(\begin{array}{cc}
1 & \frac{P(f) \gamma(f)}{1+P(f)} \\
\frac{P(f) \gamma(f)}{1+P(f)} & 1
\end{array}\right),
\end{aligned}
\label{eq:onesided_psd_matrix}
\end{equation}
in the equation, we ignore $\frac{1}{2} \delta\left(f-f^{\prime}\right)$ for simplicity, and we define $P(f) \equiv S_{C}(f)/S_{D}(f)$ as the PSD ratio between the confusion noise and the instrumental noise. As we can see, if there is no confusion noise ($P(f)=0$), the PSD matrix becomes a diagonal matrix, the diagonal element is just the PSD of the detector noise. But in general cases, the diagonal element is the approximation for the PSD of the total noise $S_{h}^{tot}(f) \approx S_{D}(f)+S_{C}(f)$. We use ``$\approx$'' because we ignore the phase difference between $S_{D}(f)$ and $S_{C}(f)$, but the ensemble average can approximately eliminate this random phase difference.

\begin{figure}[h]
	\includegraphics[width=\linewidth,scale=0.6]{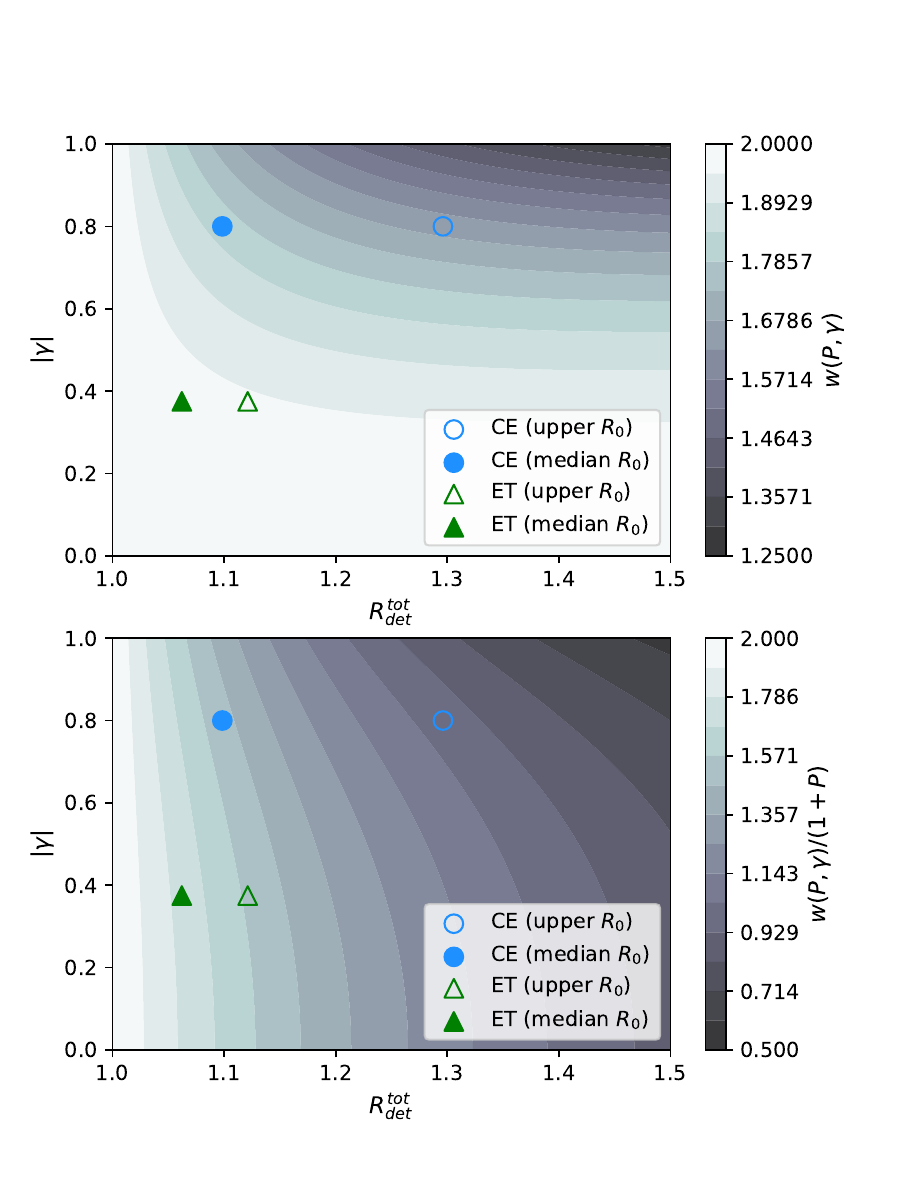}
    \caption{The upper plot shows the value of the weight $w(f,P,\gamma)$ defined in Eq.~(\ref{eq:weight}), the lower one is the weight $w(f,P,\gamma)$ divided by $1 + P(f)$ [For their specific meaning, see the text after Eq.~(\ref{eq:optimal_snr_sky_averaged_ratio})]. The $\left | \gamma \right |$ is the absolute value of the overlap reduction function (ORF) between two GW detectors. The $R_{det}^{tot}$ is the ASD ratio between the total noise and the instrumental noise, note that $P(f) \approx \left[R_{det}^{tot}(f) \right]^{2} - 1$, which is the PSD ratio between the confusion noise and the instrumental noise. When the weight $w(f,P,\gamma)$ is very close or equal to 2, that means the quadrature summation rule of network SNR [Eq.~(\ref{eq:network_snr})] is still valid. We use circles (triangles) to mark the most extreme weights of CE (ET) network for the median rate case (solid) and upper rate case (hollow). These maximum $R_{det}^{tot}$ values are selected from the median and upper merger rate cases in Fig.~\ref{fig:asd_bias_and_ratio_CE} and \ref{fig:asd_bias_and_ratio_ETD}, then we use the corresponding frequency to select $\left | \gamma \right |$.}
    \label{fig:correlated_noise_network_snr_loss_all}
\end{figure}

Following \citep{Cutler:1994ys,Seto:2021crt}, we generalize the discussion and conclusions with the sky-averaged SNR, rather than for a CBC signal with specific parameters. We replace the numerator in Eq.~(\ref{eq:optimal_snr}) with the sky-averaged $\left\langle h_{\mathrm{I}}(f) h_{\mathrm{I}}(f)^{*}\right\rangle_{\beta}$, here we use the ``$\beta$'' to abstractly represent sky localization and polarization angle, ``$\langle\cdot\cdot\rangle_{\beta}$'' means average over these parameters. We additionally account for the variation of each term with the frequency and the integral over the frequency range,
\begin{equation}
\sigma_{1 \beta}^{2}=4 \int_{f_{\text {min}}}^{f_{\text {max}}} \frac{\left\langle h_{\mathrm{I}}(f) h_{\mathrm{I}}(f)^{*}\right\rangle_{\beta}}{S_{D}(f)+S_{C}(f)} \mathrm{~d} f,
\label{eq:optimal_snr_sky_averaged}
\end{equation}
similar to the noise matrix, we have
\begin{equation}
\begin{aligned}
&\left\langle h_{\mathrm{I}}(f) h_{\mathrm{I}}(f)^{*}\right\rangle_{\beta}=\left\langle h_{\mathrm{II}}(f) h_{\mathrm{II}}(f)^{*}\right\rangle_{\beta}, \\
&\left\langle h_{\mathrm{I}}(f) h_{\mathrm{II}}(f)^{*}\right\rangle_{\beta}=\left\langle h_{\mathrm{II}}(f) h_{\mathrm{I}}(f)^{*}\right\rangle_{\beta}=\gamma(f)\left\langle h_{\mathrm{I}}(f) h_{\mathrm{I}}(f)^{*}\right\rangle_{\beta},
\end{aligned}
\label{eq:optimal_snr_sky_averaged_matrix_element}
\end{equation}
so we have the signal matrix for this two-detector network
\begin{equation}
\begin{aligned}
H_{m n}(f) & =\left\langle h_{m}(f) h_{n}(f)^{*}\right\rangle_{\beta} \\
&=\left\langle h_{\mathrm{I}}(f) h_{\mathrm{I}}(f)^{*}\right\rangle_{\beta}\left(\begin{array}{cc}
1 & \gamma(f) \\
\gamma(f) & 1
\end{array}\right),
\end{aligned}
\label{eq:signal_matrix}
\end{equation}
multiplied by the inverse of the PSD matrix [Eq.~(\ref{eq:onesided_psd_matrix})], we can generalize Eq.~(\ref{eq:optimal_snr_sky_averaged}) to $\sigma_{2 \beta}^{2}=4 \int_{f_{\text {min}}}^{f_{\text {max}}} \operatorname{tr}\left[H_{m n}(f) S_{m n}(f)^{-1} \right ] \mathrm{~d} f$, the off-diagonal elements in the matrix $H_{m n}(f) S_{m n}(f)^{-1}$ are caused by the correlated noise. If we divide it by $\sigma_{1 \beta}^{2}$, we get
\begin{equation}
\begin{aligned}
\frac{\sigma_{2 \beta}^{2}}{\sigma_{1 \beta}^{2}}
&=\frac{\int_{f_{\text {min}}}^{f_{\text {max}}} \frac{2\left[1+P(f)\right]\left[1+P(f)-\gamma(f)^{2} P(f)\right]}{\left[1+P(f)\right]^{2}-P(f)^{2} \gamma(f)^{2}} \frac{\left\langle h_{\mathrm{I}}(f) h_{\mathrm{I}}(f)^{*}\right\rangle_{\beta}}{S_{D}(f)+S_{C}(f)} \mathrm{~d} f}{\int_{f_{\text {min}}}^{f_{\text {max}}} \frac{\left\langle h_{\mathrm{I}}(f) h_{\mathrm{I}}(f)^{*}\right\rangle_{\beta}}{S_{D}(f)+S_{C}(f)} \mathrm{~d} f},
\end{aligned}
\label{eq:optimal_snr_sky_averaged_ratio}
\end{equation}
and we define the weight $w(f,P,\gamma)$ as
\begin{equation}
w(f,P,\gamma) \equiv \frac{2\left[1+P(f)\right]\left[1+P(f)-\gamma(f)^{2} P(f)\right]}{\left[1+P(f)\right]^{2}-P(f)^{2} \gamma(f)^{2}},
\label{eq:weight}
\end{equation}
``tr'' means take the trace of a matrix. When there is no confusion noise, i.e., $P(f)=0$, then Eq.~(\ref{eq:optimal_snr_sky_averaged_ratio}) is 2; this is equivalent to the quadrature summation rule. 

In order to more easily compare the information from the ASD ratio plots, we replace $P(f) \approx \left[R_{det}^{tot}(f) \right]^{2} - 1$, where $R_{det}^{tot}(f)$ is the ASD ratio of total noise to that of the instrumental noise. Here we use ``$\approx$'' also for the ignorance of the phase. In Fig.~\ref{fig:asd_bias_and_ratio_CE} and \ref{fig:asd_bias_and_ratio_ETD} we see that the maximum of $R_{det}^{tot}(f)$ is around 1.30 (1.12) for the CE (ET) upper rate case, and around 1.10 (1.06) for the CE (ET) median rate case. We plot the weight $w(f,P,\gamma)$ in Eq.~(\ref{eq:weight}) as the upper plot in Fig.~\ref{fig:correlated_noise_network_snr_loss_all}, the lower one is the weight $w(f,P,\gamma)$ divided by $1+P(f)$, which is equivalent to $\sigma_{2 \beta}^{2}$ (with the confusion noise) divided by $\sigma_{1 \beta}^{2}$ (without the confusion noise). The upper plot shows the network SNR loss only caused by the noises' correlation, the lower one also includes the loss from the biased PSD in the single detector case. 

For E1, E2, and E3 in ET, $\left | \gamma(f) \right |$ is around 0.375 when $f$ below 100 Hz \citep{Regimbau:2012ir}, combined with the maximum ASD ratio mentioned above, we use triangles to mark the most extreme weights of ET in Fig.~\ref{fig:correlated_noise_network_snr_loss_all}. We can see that the quadrature summation rule of network SNR is still valid for ET in both rate cases. As for CE, although the final sites of two CE detectors have not yet been selected, if we assume they have the same position and orientation as the current two Advanced LIGO detectors, we can use the $\left | \gamma(f) \right |$ from \citep{Fotopoulos:2008yq}, where the $\left | \gamma(f) \right |$ can be around 0.8 at 15 Hz. The weight $w(f,P,\gamma)$ might be reduced to around 1.8 (1.6) at most for median (upper) rate case. For the $w(f,P,\gamma)/(1+P)$ of CE, it can be around 1.5 (1.0) at the most for median (upper) rate case.

If a signal has a low detector-frame total mass (such as a nearby BNS signal), then it will cross a wide range of frequencies; after the integration in Eq.~(\ref{eq:optimal_snr_sky_averaged_ratio}), the final ratio should still be very close to 2. However, for a signal with a very high detector-frame total mass (such as the high redshift signal or the IMBH signal), the frequency range within the detector is very narrow, then the network SNR loss due to the correlated confusion noise will not be negligible. But in general, we can still assume the quadrature summation rule of network SNR is approximately true for two CE detectors, especially so if mitigation is applied as described in the next section.

\section{\label{sec:method}Subtraction of Binary Neutron Star Signals}

As shown in Sec.~\ref{sec:biases_from_confusion}, the large population of overlapping signals will have a negative impact on the sensitivity of the 3G ground-based GW detectors; this is primarily due to biases in the estimation of the PSD. One straightforward method to reduce the impact is to subtract the known signals from the data. It can be seen from Sec.~\ref{sec:bias_from_psd} that BNS signals have the greatest impact on 3G detectors, so for simplicity, in this section, we will focus on the impact of a BNS-only population. Here we choose to do only a single-detector search and subtraction, which is conservative if multiple detectors are operating in a network, however, it demonstrates the worst scenario that only a single highly sensitive detector is operating.

Previous papers have used similar operations in the detection of the cosmological stochastic GW background. For example, \citep{Cutler:2005qq} and \citep{Harms:2008xv} discuss how to reduce the BNS foreground signals from the Big Bang Observer (BBO) data, and then use the ``residual noise projection” method to further reduce the residual noise in the previous subtraction step, making it possible to detect the cosmological stochastic GW background through the standard cross-correlation method. There are also related studies on signal subtraction to detect the cosmological stochastic GW background of the 3G ground-based GW detectors, such as \citep{Sharma:2020btq} and \citep{Sachdev:2020bkk}. According to Sec.~\ref{sec:biases_from_confusion}, signal subtraction may also be required for more typical CBC searches. As a first step, we test a straightforward signal subtraction method that can be performed without detailed knowledge of the foreground signal population and estimates of their source parameters.

In Sec.~\ref{sec:biases_from_confusion}, we described the basic principle of matched filtering. When conducting a search, while we may maximize over the extrinsic parameters analytically as mentioned in Sec.~\ref{sec:biases_from_confusion}, to maximize over the intrinsic parameters, a bank of waveform templates is used which is designed to cover the target parameter space of the analysis \citep{Allen:2005fk}. The goal is to find a set of discrete lattice points in the intrinsic parameter space such that any point in this parameter space matches a template in the bank with a degree higher than a certain threshold (usually the minimal match is 0.97). This ensures that the number of missed signals due to the bank's discreteness can be minimized. At present, the methods of generating template banks can be divided into the stochastic \citep{Babak:2008rb,Messenger:2008ta,Harry:2009ea,Manca:2009xw,Allen:2022lqr}, geometric \citep{Owen:1995tm,Cokelaer:2007kx,Roulet:2019hzy}, and hybrid \citep{Roy:2017oul,DalCanton:2017ala,Roy:2017qgg,Coogan:2022qxs}. 

In this paper, we use \texttt{PyCBC}'s stochastic method used in the \texttt{4-OGC} search \citep{Nitz:2021zwj}. Since the low-frequency cutoff of ET and CE will be as low as 2 Hz and 5 Hz, therefore, the BNS signal will last for hours or even days in the frequency band of the detector \citep{Zhu:2021ram}. For the signal subtraction, we choose to generate a template bank starting only from 10 Hz, where the signals will be shorter than an hour. It would be expected that this initial analysis would be followed by a deep analysis after the initial
subtraction has been performed. We use \texttt{IMRPhenomD} to generate the template bank, the reference frequency $f_{ref}$ and low-frequency cutoff $f_{min}$ are both 10 Hz, and the sampling rate $f_{s}$ is 4096 Hz. We consider the case of no spin, so the intrinsic parameter is the detector-frame total mass and mass ratio. According to the simulation in Sec.~\ref{sec:population_model} and \ref{sec:bias_from_psd}, 3G detectors might detect the BNS from high redshift, so the detector-frame total mass might be several tens of times solar masses. For bank generation, we choose $[2.4, 60]$ $M_{\odot}$ as the detector-frame total mass range, and $[1, 1.636]$ as the mass ratio range. The lower bound of the total mass is consistent with the two lowest detector-frame mass NSs in our population simulation, and the upper bound of total mass covers BNS from high redshift. The upper bound on mass ratio is calculated by the lowest and highest mass NS from our population model. Two template banks (with a minimal match of 0.97) are generated using the design sensitivities of ET and CE, requiring 95652 and 152537 templates for CE and ET, respectively.

 We analyze mock data containing only BNS signals but otherwise following Sec.~\ref{sec:confusion_gen}. There are two sets of data for CE, corresponding to the median local merger rate dataset and the upper local merger rate dataset. Similarly, there are also two datasets for ET. We divide each dataset into several 1-hour data segments with 50\% overlap, and use the Welch method to estimate the PSD of the first data segment for matched filtering. For each template, we calculate the corresponding
SNR time series and record the filter's value, time, and template parameters for each SNR sample that exceeds $\left| \rho \right|=6$. After the entire template bank is searched, we need to cluster the triggers because the SNR time series peak of the trigger itself has a width (as can be seen from Fig.~\ref{fig:snr_timeseries_rayleigh_CE}), and different templates generate triggers for the same injected signal. We apply a sliding time window of 1 s and select the trigger with the largest $\left| \rho \right|$ within this window. For this study, we neglect triggers that 
correspond to signals (from 10 Hz) which are only partially within the dataset.

\begin{figure}[h]
	\includegraphics[width=\linewidth,scale=0.6]{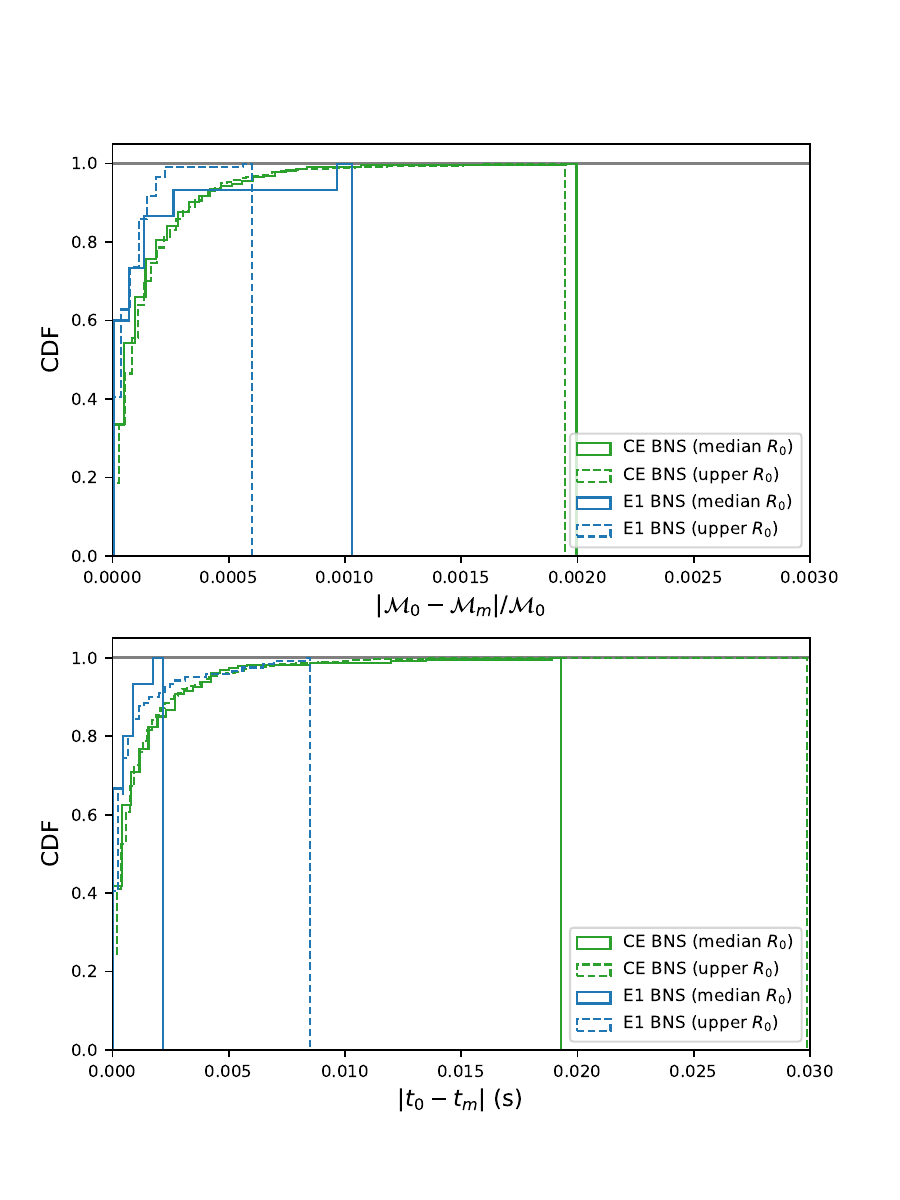}
    \caption{The parameter accuracy of the detector-frame chirp mass and coalescence time recovered by the simplified single-detector matched filtering search. The upper plot shows the cumulative distribution function for deviation of the detector-frame chirp mass of all true alarms in each dataset. The bottom plot shows the accuracy of coalescence time. More than 90\% of true alarms in each dataset has a deviation of detector-frame chirp mass smaller than 0.1\% and a deviation of merger time smaller than 0.005 s. Note that E1 means the first subdetector in ET.}
    \label{fig:pe_bias_cdf_all}
\end{figure}

The number of false alarms caused by the detector's stationary and Gaussian noise is related to the number of templates $N_{t}$ in the template bank (strictly, we should use the number of effective templates, because some templates in the stochastic bank are redundant), the sampling rate $f_{s}$, the data duration $T$, and the SNR search threshold $\rho_{th}$, then the number of false alarms can be roughly estimated by $N_{t} f_{s} T e^{-\rho_{th}^{2}/2}$.

In our search, at the SNR threshold of 6, most of the triggers are false alarms, however, we find that at a threshold of 7, the triggers are mostly true alarms, this behavior is also predicted by the equation in the last paragraph. At a threshold of 7 and for CE's median local merger rate BNS dataset, the number of true alarms is about 40\% of the total injected signals. For CE's upper local merger rate BNS dataset, the number of true alarms is about 33\% of all injections. In contrast, for the two BNS datasets of ET (strictly speaking, we just use the E1 detector, not include E2 and E3), the corresponding numbers are just around 3\%, because the majority of injected signals for ET have the optimal SNR lower than 7, note that the $f_{min}$ used in our search also limits the capability of ET.

To illustrate the parameter accuracy of these true alarms (parameters obtained by the matched filtering can be regarded as point estimates of true parameters), we show the accuracy of the detector-frame chirp mass and the merger time in Fig.~\ref{fig:pe_bias_cdf_all}, $\mathcal{M}_{0}$ and $t_{0}$ are the true detector-frame chirp mass and true merger time of the injected signal, respectively, while $\mathcal{M}_{m}$ and $t_{m}$ are the point estimates obtained by the matched filtering. When we compare triggers and injections, we consider the detector-frame chirp mass and the merger time to be the same up to one decimal place as true alarms. We use the cumulative distribution function (CDF) to show the bias of these point estimates. More than 90\% of true alarms in all datasets have a deviation of chirp mass smaller than 0.1\% and a deviation of merger time smaller than 0.005 s. For the CE upper rate case, it has a longer tail to larger deviations, this is caused by the possibly very close signals \citep{Samajdar:2021egv,Pizzati:2021apa,Smith:2021bqc,Relton:2021cax,Himemoto:2021ukb,Antonelli:2021vwg} or the misassociation issue, sometimes there might be more than one trigger that meets our true alarm criteria (within 0.1 s and 0.1 $M_{\odot}$) when compared with injections, so we might choose a nearby but wrong injection.

\begin{figure*}
    \centering
	\includegraphics[scale=0.6]{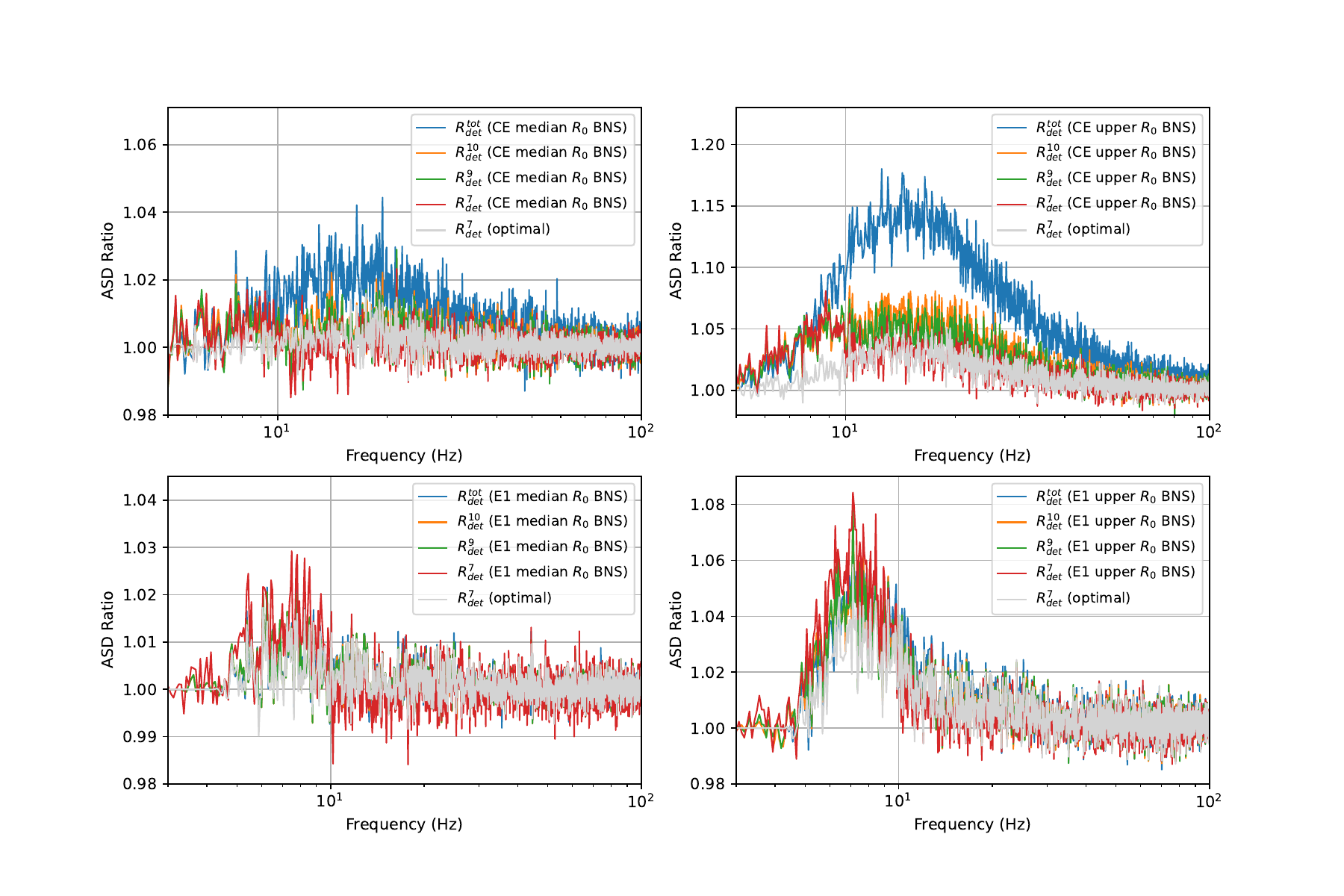}
    \caption{The ratio of the ASD after signal subtraction at different SNR thresholds to the detector noise ASD for CE (upper panels) and E1 (lower panels) and both the median (left panels) and upper merger rate scenarios (right panels). The blue lines represent the ASD deviation before the signal subtraction. The orange, green, and red lines represent the results after subtracting the rescaled trigger waveforms with SNR thresholds of 10, 9, and 7, respectively. The gray lines represent the optimal subtraction of all signals with an optimal SNR greater than 7. For CE, the deviations peak around 15 Hz, our best subtraction results can almost achieve the optimal results above 10 Hz. For E1, the deviations peak around 7 Hz, we cannot achieve better ASD by using the current subtraction method, because most signals are below the SNR threshold 7.}
    \label{fig:asd_ratio_subtraction}
\end{figure*}

For each identified candidate trigger, we rescale the corresponding template according to the trigger's complex SNR, to reconstruct an estimate of the injected signal.
If we express the complex SNR we obtained through the matched filtering as $\left| \rho_{m} \right| e^{i\varphi_{m}}$ and the template associated with the trigger is $h_{+}$,
we get the rescaled template with the following formula
\begin{equation}
\widehat{h}(t,\Theta_{m}) = \frac{\left| \rho_{m} \right|}{\sigma_{f_{min}^{I}}^{\frac{1}{2}}} 
\int_{f_{min}^{II}}^{f_{max}}\widetilde{h}_{+,f_{min}^{II}}(f,\Theta_{m})e^{i\left[2\pi f(t+t_{m}) + \varphi_{m} \right]}df,
\label{eq:rescaled_template}
\end{equation}
here $\Theta_{m}$ means all the intrinsic parameters measured by matched filtering. The $f_{min}^{I}$ is the $f_{min}$ used in the matched filtering search, in our case, it is 10 Hz. The $\sigma_{f_{min}^{I}}^{\frac{1}{2}}$ is the optimal SNR of the trigger's template used in matched filtering. The $f_{min}^{II}$ is the $f_{min}$ used to generate the signal in datasets and subtraction, in our case, 5 Hz for CE and 2 Hz for ET.

We sequentially subtract the rescaled template $\widehat{h}(t,\Theta_{m})$ corresponding to all triggers from the dataset. As mentioned before, the number of false alarms is related to the threshold  $\left| \rho_{th} \right|$. If the threshold is too low (so the match between the template and data is not good enough, the SNR is just contributed by a small portion of the template), it means that we are almost injecting the antiphased waveform of false alarms that were not originally in the dataset, which will increase the deviation of the PSD instead. So in order to examine the effect of different thresholds on the signal subtraction, we select $\left| \rho_{th} \right|\in\left \{ 7,8,9,10 \right \}$ to filter triggers, if the threshold is 6, we find that too many false alarms will make the deviation higher than the case without subtraction, so we do not show the results with a threshold of 6. The ASD ratios after the signal subtraction are shown in Fig.~\ref{fig:asd_ratio_subtraction}, we do not show the results for $\left| \rho_{th} \right| = 8$ to make the plot clearer, because they are just between results of 7 and 9. For CE's median local merger rate dataset, the maximum deviation before subtraction reaches 2\%-3\%. After subtraction by different thresholds, the maximum deviation is reduced. The results of different thresholds are generally similar, but it can still be seen that the lower the threshold, the better the subtraction; the deviation of the PSD is nearly eliminated if using a single-detector SNR threshold of 7. For the CE upper local merger rate dataset, the maximum deviation before subtraction is about 15\%. After subtracting triggers with $\left| \rho_{th} \right|$ above 10, it is significantly reduced to 6\%. It can be seen that the deviation of PSD mainly comes from high SNR signals ($\left| \rho \right|$ above 10). Similar to the median rate dataset, as the threshold decreases, the ASD ratio reduces overall, and the best result is the red line. For the ET median rate cases, the bias (about 1\%-2\%) is around 7 Hz, and the signal subtraction cannot reduce the ASD bias. For the ET upper rate cases, the results are similar to median rate cases, but with a higher bias peak of around 7 Hz.

In order to understand the results of signal subtraction, we also show the results obtained by excluding all signals with the optimal SNR higher than 7 in the corresponding dataset, which we can regard as an ``optimal subtraction'' when $\left| \rho_{th} \right|=7$, because these results are not affected by false alarms, and there is no residual noise caused by template parameters' deviation. For the results of the CE median rate, the achieved subtraction and the optimal subtraction are consistent above 10 Hz, and in the 5-10 Hz interval, however, the subtraction is marginally worse than the optimal reference, this is because we just extrapolating the waveform to this frequency range [see Eq.~(\ref{eq:rescaled_template})], that means much larger mismatch and residual noise here. For the CE upper rate results, similarly, in the frequency range above 10 Hz, there is an agreement between our subtraction and the optimal reference, however, in the 5-10 Hz interval, no significant subtraction is observed. For the ET median (upper) rate cases, all results are very similar to the optimal reference, which means the remaining bias is mainly due to signals with the optimal SNR lower than 7.

\begin{figure*}
    \centering
	\includegraphics[scale=0.6]{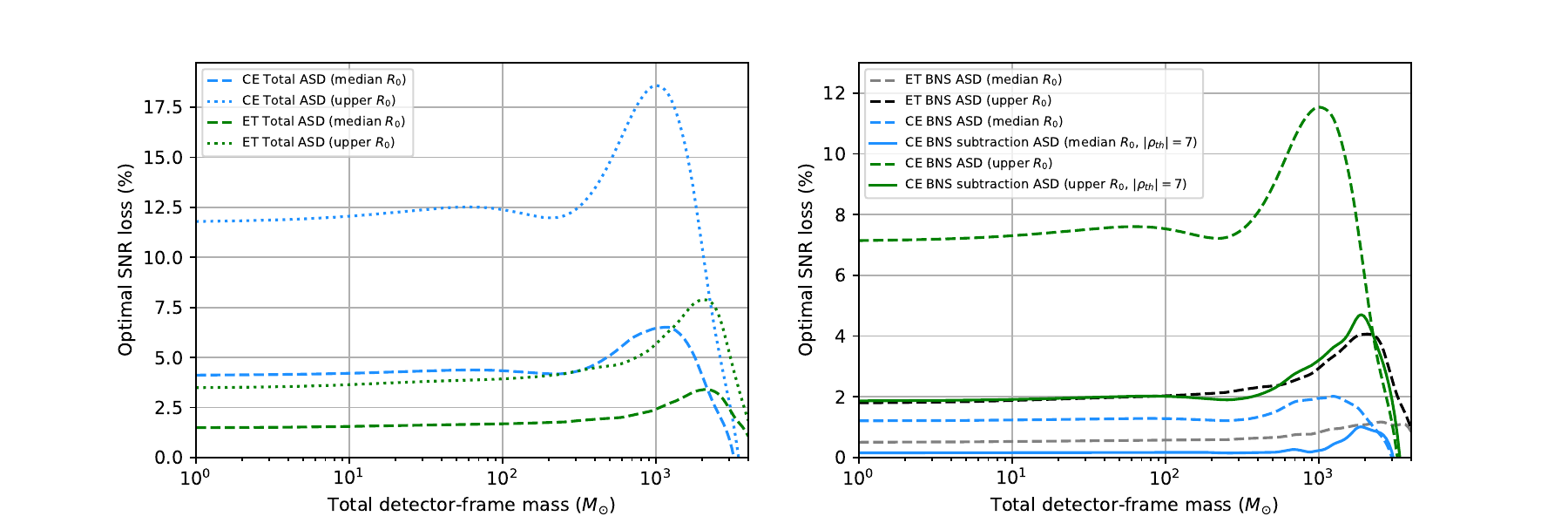}
    \caption{The optimal SNR loss as a function of total detector-frame masses. The left plot shows the loss when accounting for all source types, while the right plot shows the results before and after the signal subtraction, but for data including only BNS signals. In general, the larger the detector-frame total mass (less than 1000-2000 $M_{\odot}$), the higher the SNR loss, then it drops quickly to almost zero; for 1000-2000 $M_{\odot}$ systems, the majority of the signal is contained within the most biased frequency ranges. The SNR loss peaks at higher masses for ET than CE because the PSD bias is shifted to lower frequencies. By comparing the left and right plots, it can be seen that the main cause of SNR loss is the ASD deviation caused by BNS signals. The SNR loss in CE caused by confusion noise is more significant than in ET due to its higher BNS sensitivity.
    }
    \label{fig:optsnr_loss_CE_ET_detframe_totM_IMRPhenomD_all}
\end{figure*}

To quantitatively study the improvement in detection ability brought by the straightforward signal subtraction, we calculate the optimal SNR loss (assuming the optimal orientation) of equal mass CBC signals with different detector-frame total mass under different ASDs, as shown in Fig.~\ref{fig:optsnr_loss_CE_ET_detframe_totM_IMRPhenomD_all}. It can be seen that the overall effect of confusion noise on ET is lower than CE. Even for the upper rate total ASD, the observed SNR loss in ET is around 4\%. For the median rate total ASD, the overall loss is lower than 2.5\%. In contrast, the loss of CE's upper rate total ASD is around 12\%. For a signal with a total detector-frame mass of 1000 $M_{\odot}$, the loss will even reach 18\%, and for the median rate total ASD, the loss is also around 5\%. For CE, we show the results before and after BNS signal subtraction ($\left| \rho_{th} \right|=7$), for the median rate BNS ASD, after signal subtraction, the overall loss can be reduced from 1.2\% to around 0.2\%. For the upper rate BNS ASD, the loss can be greatly reduced from around 8\% to 2\%. These results are consistent with Fig.~\ref{fig:horizon_redshift_loss_ratio_CE_ET_sourceframe_totM_IMRPhenomD}. For ET, it can be seen from Fig.~\ref{fig:asd_ratio_subtraction} that very few BNS signals can be detected and subtracted from only ET's subdetector E1. Even for the upper rate BNS ASD, the loss of ET is below 5\% for most sources.

Here we discuss if we can further improve the signal subtraction results. As we have mentioned before, the residual noise and the SNR threshold (or false alarms) limit the capability of signal subtraction. First, let us discuss the origin of the subtraction residual. If we regard the template bank as a template manifold \citep{Cutler:1994ys}, the parameters of the real signal are located in $h(t,\Theta)$ in the manifold if there is no systematic error in the waveform modeling. Due to the existence of the detector noise $n$, the total signal $s$ measured by the detector (here we ignore overlapping signals for simplicity) will be located outside the template manifold. Using the entire template bank to perform matched filtering on the data $s$ is equivalent to finding the closest point $\widehat{h}(t,\Theta_{m})$ to $s$ in the template manifold, so the vector between point $s$ and point $\widehat{h}(t,\Theta_{m})$ should be perpendicular to the tangent plane at $\widehat{h}(t,\Theta_{m})$ in the manifold (see Fig.~1 in \citep{Cutler:1994ys}). In fact, the inner product~\ref{eq:inner_product} can be regarded as the projection of $a$ on $b$, then we have
\begin{equation}
\begin{array}{ll}
&\left \langle s - \widehat{h}(t,\Theta_{m}) \mid \partial_{\Theta_{m}}  \widehat{h}(t,\Theta_{m})  \right \rangle \\
=& \left \langle n_{\perp} + n_{\parallel} + h(t,\Theta) - \widehat{h}(t,\Theta_{m}) \mid \partial_{\Theta_{m}}  \widehat{h}(t,\Theta_{m}) \right \rangle \\
=& \left \langle n_{\perp} + n_{\parallel} - (\widehat{h}(t,\Theta_{m}) - h(t,\Theta)) \mid \partial_{\Theta_{m}}  \widehat{h}(t,\Theta_{m}) \right \rangle = 0,
\label{eq:residual_manifold}
\end{array}
\end{equation}
among them $-r(t) = \widehat{h}(t,\Theta_{m}) - h(t,\Theta)$ is the residual noise (with the opposite sign). Here we decompose the detector noise $n$ into $n_{\perp}$ and $n_{\parallel}$, which are the perpendicular and parallel components at the point $\widehat{h}(t,\Theta_{m})$, respectively. As we can see here, $\left \langle n_{\perp} \mid \partial_{\Theta_{m}}  \widehat{h}(t,\Theta_{m}) \right \rangle$ should be 0, so $n_{\parallel} = \widehat{h}(t,\Theta_{m}) - h(t,\Theta) = -r(t)$, which means the residual noise is caused by the parallel component of detector noise $n$ at the point $\widehat{h}(t,\Theta_{m})$. The residual noise can be further reduced by the residual noise projection method after the first-stage signal subtraction \citep{Cutler:2005qq,Harms:2008xv}. This follow-up method is based on the Fisher information matrix (FIM) and the signal manifold \citep{Cutler:1994ys}, which requires the signal to have a sufficiently high SNR, that is, satisfying the linear signal approximation (LSA) \citep{Finn:1992wt}, which is not valid for many low SNR signals in our simulations. As we can see in Fig.~\ref{fig:asd_ratio_subtraction}, above 10 Hz, even without this follow-up step, the ASD biases are already minimal for 3G detectors. However, it might be worth investigating if we can further remove some part of the residual noise below 10 Hz by using a similar second-stage method (without lowering the $f_{min}$ = 10 Hz in our bank generation and search); for the SNR threshold or false alarms, according to the results from Sec.~\ref{sec:bias_from_correlated_noises}, the quadrature summation rule of the network SNR still approximately holds for the 3G network, so we can utilize multiple detectors to increase the total SNR and lower the SNR threshold in each detector, that means we can detect and subtract more signals. Also, we can use the coincidence test of the arrival time and the consistent test of the template’s parameters between different detectors, to reduce the number of false alarms \citep{Usman:2015kfa}.

\section{\label{sec:conclusions}Conclusions}

In this paper, we simulated the time-domain strain data of the 3G ground-based GW detectors CE and ET based on the latest \texttt{GWTC-3} population results (see Fig.~\ref{fig:median_merger_rate} and \ref{fig:strain_median_CE}). Due to the improved low-frequency sensitivity and much higher detection rate of the 3G ground-based GW detectors, GW signals will overlap each other, forming confusion noise.
Since the matched filter is the optimal linear filter only under stationary and Gaussian noise, the addition of the correlated non-Gaussian confusion noise (see Fig.~\ref{fig:snr_timeseries_rayleigh_CE}) might have an impact on the performance of matched filtering-based searches. We quantitatively investigated the factors that might affect the performance of matched filtering, such as the deviation caused by confusion noise to the ASD, the deviation caused by cross terms in the inner product, and the correlation of confusion noise among the detector network, which might break the standard quadrature summation rule of the network SNR. 

We found that the most significant impact from confusion noise on matched filtering comes from biases in PSD or ASD estimation (see Fig.~\ref{fig:asd_bias_and_ratio_CE} and \ref{fig:asd_bias_and_ratio_ETD}). We used the horizon distance (redshift) for different sources to measure the loss caused by the biased ASD (see Fig.~\ref{fig:horizon_redshift_loss_ratio_CE_ET_sourceframe_totM_IMRPhenomD}). For ET, the confusion noise made by median (upper) local merger rate estimates of CBC sources will reduce the horizon redshift by up to 8 (21)\%. For CE, the deviation of the ASD is much larger; the horizon redshift can be reduced by 15 (38) \% for the median (upper) merger rate scenarios. A portion of PBH and Pop III sources from redshift higher than 20 may be missed if the impact of confusion noise is left unmitigated; these GW sources are important scientific targets for future 3G detectors~\cite{evans2021horizon}. In addition, subsolar compact binaries and high-redshift BNSs will also be affected. The population of sources whose total source-frame mass is higher than 100 $M_{\odot}$ will still be fully detected but with reduced SNR for nearby signals.

In addition to a biased ASD, the presence of a foreground population of signals could directly contribute to the matched filter if there is an overlap between sources. As expected, we found that confusion noise has different properties than the detector's stationary and Gaussian noise (see Fig.~\ref{fig:snr_timeseries_rayleigh_CE}), but in general its contribution to the SNR is significantly smaller than the instrumental noise. If the merger times of adjacent CBC signals are close enough, however, the contribution of cross terms between sources can no longer be ignored; the effect on parameter estimation has been studied in numerous works \citep{Samajdar:2021egv,Pizzati:2021apa,Smith:2021bqc,Relton:2021cax,Himemoto:2021ukb,Antonelli:2021vwg}. 

We also investigated the SNR loss caused by the noises' correlation in 3G detector networks. We adopted the method from \citep{Seto:2021crt} and used the effective number of detectors as a function of the overlap reduction function and ASD ratio to quantify this loss. Combined with the results of our biased ASD, we found that the quadrature summation rule of the network SNR is still approximately valid for ET, but might be modified for high detector-frame mass signals for a network of two CE detectors (see Fig.~\ref{fig:correlated_noise_network_snr_loss_all}).

In order to reduce the influence of confusion noise on the ASD, we tested a straightforward single-detector signal subtraction (see Fig.~\ref{fig:asd_ratio_subtraction}) that can be implemented at a minimal computational cost relative to a full search. We examined our method with different SNR thresholds for BNS datasets with different local merger rates. BNS signals are the main contributor to the confusion noise, especially for upper local merger rate cases, and it is straightforward to extend our method to NSBH cases. For CE, when the SNR threshold is 7, we obtained nearly optimal subtraction results, almost back to the instrumental noise level. Since the vast majority of signals in E1 are lower than our minimum threshold, the current signal subtraction results of ET are not ideal; the null stream method of ET is needed as a supplement to our method \citep{Regimbau:2012ir,Meacher:2015rex}. For CE, our method can limit the SNR loss to 0.2\% (median BNS rate) and 2\% (upper BNS rate) in general (see Fig.~\ref{fig:optsnr_loss_CE_ET_detframe_totM_IMRPhenomD_all}). Our demonstrated signal subtraction procedure can be used as a first-stage foreground cleaning, allowing for more sophisticated follow-up stages. Our results show that this straightforward single-detector implementation is sufficient to enable the archival detection of typical binary mergers. For the early warning of mergers with 3G detectors \citep{Nitz:2021pbr}, we might expect more significant biases for high-redshift mergers, however, expect the detection of nearby, optically bright, sources would not be significantly impacted. 

Our current signal extrapolation and subtraction method has several constraints: (1) if the threshold of SNR is too low, there will be a large number of false alarms. Since there must be some signals below the SNR threshold, this will limit our method's capability, and (2) the $f_{min}$ used in the bank generation and signal search will affect the extrapolation accuracy of the rescaled template at frequencies lower than $f_{min}$, thereby the residual noise lower than $f_{min}$ after the subtraction is larger. Because the template bank is generated above $f_{min}$, the max mismatch of the bank (3\%) is only valid above this frequency, lower frequencies need a denser template bank at an increased computational cost. One may further lower the SNR threshold by using observations in a detector network; multiple detectors increase the total network SNR and allow for coincidence tests to reduce the contamination from false positives. It may also be worth investigating how to combine the residual noise projection method of \citep{Cutler:2005qq,Harms:2008xv} to further reduce the subtraction residual.

The code used in this research is public at \url{https://github.com/gwastro/confusion-noise-3g}. Our code is based on the \texttt{PyCBC} \citep{Usman:2015kfa}, \texttt{Python} \citep{van1995python}, \texttt{NumPy} \citep{Harris:2020xlr}, \texttt{SymPy} \citep{Meurer:2017yhf}, \texttt{SciPy} \citep{Virtanen:2019joe}, and \texttt{Matplotlib} \citep{Hunter:2007}.

\begin{acknowledgments}
We thank Jin-Ping Zhu and He Wang for the technical discussions in the early stage. We would like to acknowledge Rahul Dhurkunde and Xisco Jiménez Forteza for reading the manuscript and for providing useful comments. We are grateful to the computing team from AEI Hannover for their significant technical support.  
\end{acknowledgments}

\bibliography{apssamp}

\end{document}